\documentclass[aps,prl,twocolumn,longbibliography,superscriptaddress,notitlepage,floatfix]{revtex4-2}
\renewcommand{\figurename}{Fig.}
\usepackage{amsmath,amsfonts,amssymb,color,epsfig,graphics,graphicx,latexsym,theorem,url,multirow,diagbox}
\usepackage[colorlinks,colorlinks,citecolor=blue,linkcolor=blue,urlcolor=blue]{hyperref}
\usepackage[utf8]{inputenc}
\usepackage{booktabs}
\usepackage[T1]{fontenc}
\usepackage{courier}
\usepackage{listings}

\usepackage{tikz}
\usetikzlibrary{quantikz} 
\usepackage{ragged2e}
\usepackage{multirow}
\usepackage[thicklines]{cancel}

\usepackage{ulem}
\usepackage{latexsym}
\usepackage{dcolumn}
\usepackage{comment}
\usepackage{times}

\usepackage{algorithm}
\usepackage{algorithmic} 
\usepackage{braket}
\usepackage{bm}

\definecolor{blue}{rgb}{0,0,1}
\definecolor{red}{rgb}{1,0,0}
\definecolor{green}{rgb}{0,1,0}

\begin{document}
\title{Deep quantum neural networks equipped with backpropagation on a superconducting processor}

\author{Xiaoxuan~Pan}
\thanks{These two authors contributed equally to this work.}
\affiliation{Center for Quantum Information, Institute for Interdisciplinary Information Sciences, Tsinghua University, Beijing 100084, China}

\author {Zhide~Lu}
\thanks{These two authors contributed equally to this work.}
\affiliation{Center for Quantum Information, Institute for Interdisciplinary Information Sciences, Tsinghua University, Beijing 100084, China}

\author {Weiting~Wang}
\affiliation{Center for Quantum Information, Institute for Interdisciplinary Information Sciences, Tsinghua University, Beijing 100084, China}

\author {Ziyue~Hua}
\affiliation{Center for Quantum Information, Institute for Interdisciplinary Information Sciences, Tsinghua University, Beijing 100084, China}

\author {Yifang~Xu}
\affiliation{Center for Quantum Information, Institute for Interdisciplinary Information Sciences, Tsinghua University, Beijing 100084, China}

\author {Weikang~Li}
\affiliation{Center for Quantum Information, Institute for Interdisciplinary Information Sciences, Tsinghua University, Beijing 100084, China}

\author {Weizhou~Cai}
\affiliation{Center for Quantum Information, Institute for Interdisciplinary Information Sciences, Tsinghua University, Beijing 100084, China}

\author {Xuegang~Li}
\affiliation{Center for Quantum Information, Institute for Interdisciplinary Information Sciences, Tsinghua University, Beijing 100084, China}

\author {Haiyan~Wang}
\affiliation{Center for Quantum Information, Institute for Interdisciplinary Information Sciences, Tsinghua University, Beijing 100084, China}

\author {Yi-Pu~Song}
\affiliation{Center for Quantum Information, Institute for Interdisciplinary Information Sciences, Tsinghua University, Beijing 100084, China}

\author{Chang-Ling~Zou}
\affiliation{CAS Key Laboratory of Quantum Information, University of Science and Technology of China, Hefei, Anhui 230026, China}

\author {Dong-Ling Deng}
\thanks{E-mail: dldeng@tsinghua.edu.cn}
\affiliation{Center for Quantum Information, Institute for Interdisciplinary Information Sciences, Tsinghua University, Beijing 100084, China}
\affiliation{Shanghai Qi Zhi Institute, No. 701 Yunjin Road, Xuhui District, Shanghai 200232, China}

\author{Luyan~Sun}
\thanks{E-mail: luyansun@tsinghua.edu.cn}
\affiliation{Center for Quantum Information, Institute for Interdisciplinary Information Sciences, Tsinghua University, Beijing 100084, China}

\begin{abstract}
Deep learning and quantum computing have achieved dramatic progresses in recent years. The interplay between these two fast-growing fields gives rise to a new research frontier of quantum machine learning. 
In this work, we report the first experimental demonstration of training deep quantum neural networks via the backpropagation algorithm with a six-qubit programmable superconducting processor. In particular, we show that three-layer deep quantum neural networks can be trained efficiently to learn two-qubit quantum channels with a mean fidelity up to $96.0\%$ and the ground state energy of molecular hydrogen with an accuracy up to $93.3\%$ compared to the theoretical value. In addition, six-layer deep quantum neural networks can be trained in a similar fashion to achieve a mean fidelity up to $94.8\%$ for learning single-qubit quantum channels. 
Our experimental results explicitly showcase the advantages of deep quantum neural networks, including quantum analogue of the backpropagation algorithm and less stringent coherence-time requirement for their constituting physical qubits, thus providing a valuable guide for quantum machine learning applications with both near-term and future quantum devices.
\end{abstract}

\maketitle

\begin{figure*}[t]
\includegraphics{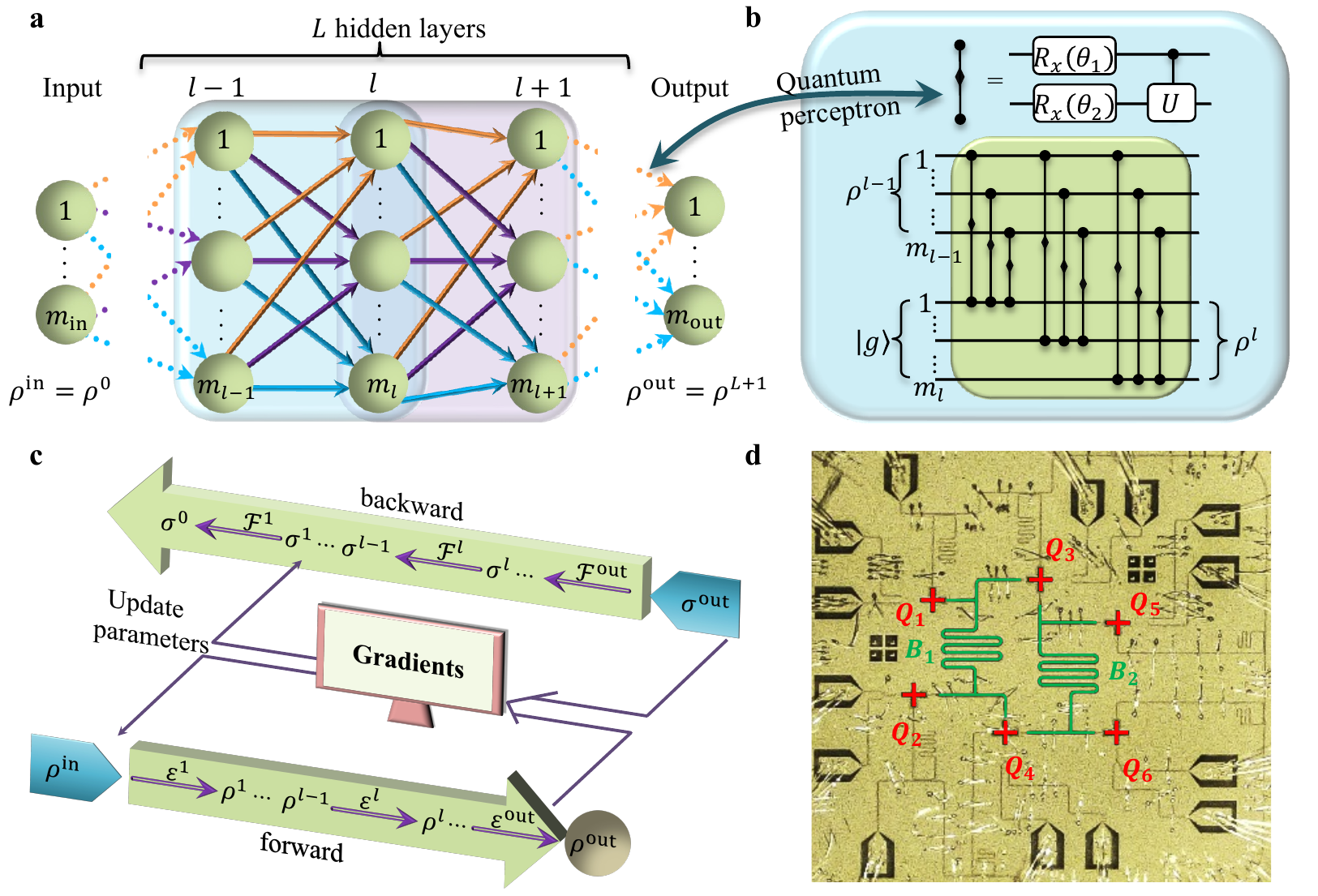}
\caption{
\textbf{A schematic of training deep quantum neural networks.}
(a), Architecture exhibition of a general DQNN. Information propagates layerwise from the input layer to the output layer. At adjacent two layers, we apply the quantum perceptron in the order according to the exhibited circuit in (b). A quantum perceptron is realized by applying two single-qubit rotation gates $R_x(\theta_1)$ and $R_x(\theta_2)$ (the rotations along the $x$ axis with variational angles $\theta_1$ and $\theta_2$, respectively) followed by a fixed two-qubit controlled-phase gate.
(c), Illustration of the quantum backpropagation algorithm. We apply forward channels $\mathcal{E}$ on $\rho^{\text {in}}$ and successively obtain $\{\rho^{1}, \rho^{2} \ldots \rho^{\text{out}}\}$, and apply backward channels $\mathcal{F}$ to successively obtain $\{\sigma^{\text{out}}, \sigma^{L} \ldots \sigma^{1}\}$ in the backward process. These forward and backward terms are used for the gradient evaluation.
(d), Exhibition of a quantum processor with six superconducting transmon qubits, which are used to experimentally implement the DQNNs. The transmon qubits ($Q_1$-$Q_6$) are marked in red and the bus resonators ($B_1$ and $B_2$) are marked in green.
}
\label{fig:framework}
\end{figure*}

Machine learning has achieved tremendous success in both commercial applications and scientific researches over the past decade. In particular, deep neural networks play a vital role in cracking some notoriously challenging problems, ranging from playing Go~\cite{Silver2017Mastering} to predicting protein structures~\cite{Jumper2021Highly}.  
They contain multiple hidden layers and are believed to be more powerful in extracting high-level features from data than traditional methods~\cite{Goodfellow2016Deep, LeCun2015Deep}. The learning process can be fueled by updating the parameters through gradient descent, where the backpropagation algorithm enables efficient calculations of gradients via the chain rule~\cite{Goodfellow2016Deep}.

By harnessing the weirdness of quantum mechanics such as superposition and entanglement, quantum machine learning approaches hold the potential to bring advantages compared with their classical counterpart.
In recent years, exciting progress has been made along this interdisciplinary direction ~\cite{Biamonte2017Quantum, Dunjko2018Machine, DasSarma2019Machine, Cerezo2022Challenges, Dawid2022Modern, Huang2021InformationTheoretic}.
For example, rigorous quantum speedups have been proved in classification models~\cite{Liu2021Rigorous} and generative models~\cite{Gao2018Quantum} with complexity-theoretic guarantees.
In terms of the expressive power for quantum neural networks, there is also preliminary evidence showing their advantages over the comparable feedforward neural networks~\cite{Abbas2021Power}.
Meanwhile, noteworthy progress has also been made on the experimental side~\cite{Herrmann2022Realizing,Ren2022Experimental, havlivcek2019supervised, Hu2019Quantum, Blank2020Quantum, Li2015Experimental, Zhu2019Training, Gong2022Quantum, Huang2022Quantum}. For examples, in Ref.~\cite{Herrmann2022Realizing}, the authors realize a quantum convolutional neural network on a superconducting quantum processor. In Ref.~\cite{Ren2022Experimental}, an experimental demonstration of quantum adversarial learning has been reported.
Similar to deep classical neural networks with multiple layers, a deep quantum neural network (DQNN) with the layer-by-layer architecture is proposed~\cite{Beer2020Training,Liu2022Solving}, which can be trained via a quantum analog of the backpropagation algorithm. 
Under this framework, the quantum analog of a perceptron is a general unitary operator acting on qubits from adjacent layers, whose parameters are updated by multiplying the corresponding updating matrix of the perceptron in the training process.

In this paper, we report the first experimental demonstration of training DQNNs through the backpropagation algorithm on a programmable  superconducting processor with six frequency-tunable transmon qubits. We find that a three-layer DQNN can be efficiently trained to learn a two-qubit target quantum channel with a mean fidelity up to $96.0\%$ and the ground state energy of molecular hydrogen with an accuracy up to $93.3\%$ compared to the theoretical prediction. In addition, we also demonstrate that a six-layer DQNN can efficiently learn a one-qubit target quantum channel with a mean fidelity up to $94.8\%$. Our approach can carry over to other DQNNs with a larger width and depth straightforwardly, thus paving a way towards large-scale quantum machine learning with potential advantages in practical applications. 

As sketched in Fig.~\ref{fig:framework}(a), our DQNN has a layer-by-layer structure, and maps the quantum information layerwise from the input layer state $\rho^{\text{in}}$, through $L$ hidden layers, to the output layer state $\rho^{\text{out}}$. Quantum perceptrons are the building blocks of the DQNN. As shown in Fig.~\ref{fig:framework}(b), a single quantum perceptron is defined as a parameterized quantum circuit applied to the corresponding qubit pair at adjacent layers, which is directly implementable in experiments.  A sequential combination of the quantum perceptrons constitutes the layerwise operation between adjacent layers. One of the key characteristics of the DQNN is the layer-by-layer quantum state mapping, allowing efficient training via the quantum backpropagation algorithm \cite{Beer2020Training}. We sketch the general experimental training process in Fig.~\ref{fig:framework}(c). When performing the quantum backpropagation algorithm, one only requires the information from adjacent two layers, rather than the full DQNN, to evaluate the gradients with respect to all parameters at these two layers. 
Such a backpropagation-equipped DQNN bears the following merit: 
it significantly reduces the requirements for the ability to maintain many coherent qubits, since qubits in each layer only need to keep their coherence for no more than the duration of two-layer operations regardless of the depth of the DQNN. This advantage makes it possible to realize DQNNs with reduced number of layers of qubits through qubit reusing~\cite{Beer2020Training}. 

\begin{figure}[t]
\includegraphics{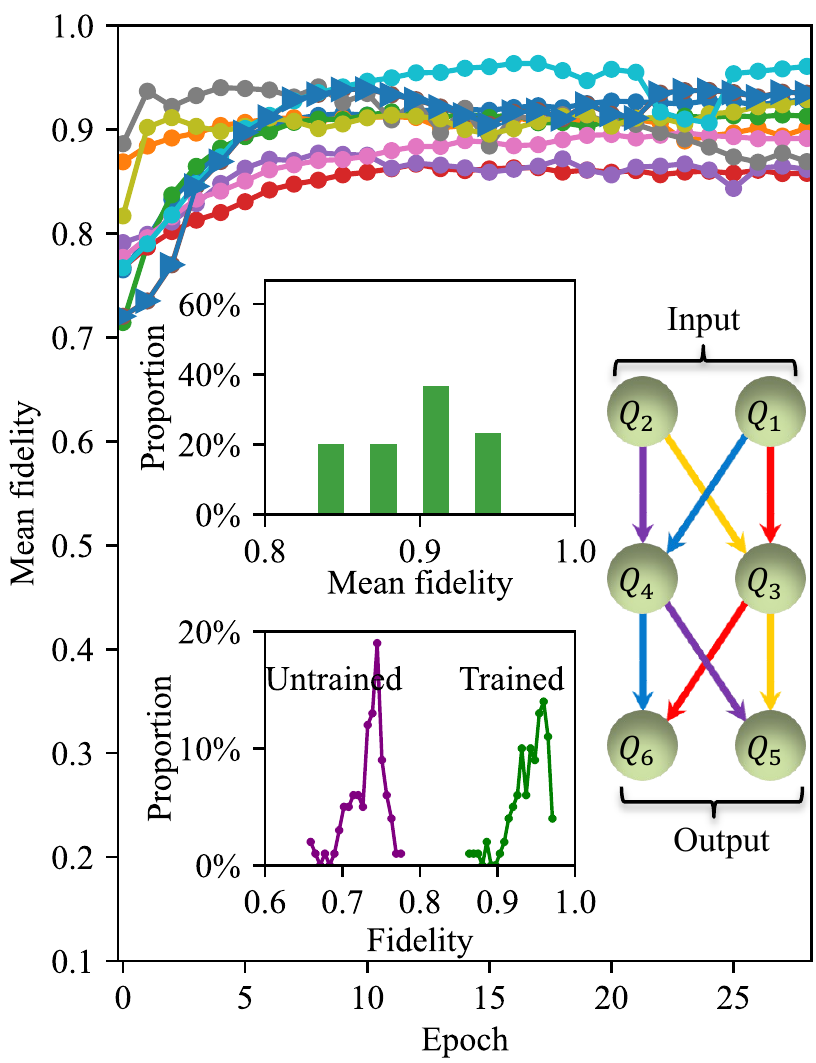}
\caption{
\textbf{Experimental results for learning a two-qubit quantum channel.}
We train the three-layer DQNN$_1$ with $30$ different initial parameters, and plot the mean fidelity as a function of training epochs for $10$ of them for clarity. The upper left inset shows the distribution of the converged mean fidelities of these $30$ different initial parameters.
We choose one of the learning curves (marked with dark blue triangles), then randomly generate $100$ different input quantum states, and test the fidelity between their output states given by the target quantum channel and the trained (untrained) DQNN$_1$. In the lower left inset, the green (purple) curve shows the distribution of the fidelities for the trained (untrained) DQNN$_1$. The right inset is a schematic illustration of DQNN$_1$. At adjacent layers, we apply the quantum perceptrons in the order indicated with the colors: red, yellow, blue, and purple.
}
\label{fig:channnel_2^3}
\end{figure}

Our experiment is carried out on a superconducting quantum processor, which possesses six two-junction and frequency-tunable transmon qubits~\cite{Koch2007Charge,Barends2013Coherent,Li2018Perfect,CaiPRL2019,Kono2020Breaking,Carusotto2020Photonic,Negirneac2021High,Blais2021Circuit}. As photographed in Fig.~\ref{fig:framework}(d), the chip is fabricated with the layout of the qubits being purposely and carefully optimized for a layer-by-layer structure. Each transmon qubit is coupled to an individual flux control line, XY control line, and quarter-wavelength readout resonator, respectively. All readout resonators are coupled to a common transmission line, which is connected through a Josephson parametric amplifier for high-fidelity single-shot readout of the qubits~\cite{Roy2015Broadband,Murch2013Observing}. In order to implement the two-qubit gates in the quantum perceptrons, two separate half-wavelength bus resonators are respectively used to mediate the interactions among the qubits between layers~\cite{Majer2007Coupling,havlivcek2019supervised,Song2019Generation}. The detailed experimental setup and device parameters can be found in Supplementary Information.

We first consider using DQNNs to learn a two-qubit quantum channel. We experimentally implement a three-layer  DQNN with two qubits in each layer. This three-layer DQNN is denoted by DQNN$_1$.
Here, we choose $\ket{00}$, $\ket{01}$, $\ket{++}$, and  $\ket{+i+i}$ as our input states $\rho_x^{\text{in}}$,  where the subscript $x=1,2,3,4$ is the labeling, $\ket{0}$ and $\ket{1}$ are the eigenstates of Pauli $Z$ matrix, $|+\rangle$ $\left(\ket{-}\right)$ is the eigenstate of Pauli $X$ matrix, and $|+i\rangle$ is the eigenstate of Pauli $Y$ matrix. The four pairs of $\left(\rho_{x}^{\text{in}},\tau_{x}^{\text {out}}\right)$ serve as the training dataset, where $\tau_{x}^{\text {out}}$ is the corresponding desired output state produced by the target quantum channel.
We learn the target quantum channel by maximizing the mean fidelity between $\tau_{x}^{\text {out}}$ and the measured DQNN output $\rho_x^\mathrm{out}$ averaged over all four input states.
The general training procedure goes as follows: 1) Initialization: we randomly choose the initial gate parameters $\boldsymbol{\theta}$ for all perceptrons in DQNN$_1$.
2) Forward process (implemented on our quantum processor): for each training sample $\left(\rho_{x}^{\text{in}},\tau_{x}^{\text {out}}\right)$, we prepare the input layer to $\rho_{x}^{\text {in}}$, then apply layerwise forward channels $\mathcal{E}^{1}$ and $\mathcal{E}^{\text {out}}$, and extract $\rho^{1}_x$ and $\rho^{\text {out}}_x$ successively by carrying out quantum state tomography \cite{Nielsen2010Quantum}.  
3) Backward process (implemented on a classical computer):  we initialize the output layer to $\sigma^{\text{out}}_x$, which is determined by $\rho_x^\mathrm{out}$ and $\tau_x^\mathrm{out}$ (see Supplementary Note 1), and then apply backward channels $\mathcal{F}^{\text{out}}$ and $\mathcal{F}^{1}$ on $\sigma^{\text{out}}_x$  to successively obtain $\sigma^{1}_x$ and $\sigma^{0}_x $.
4) Based on $\{\left(\rho_{x}^{l-1},\sigma_{x}^{l}\right)\}$, we evaluate the gradient of the fidelity with respect to all the variational parameters in the adjacent layers $l-1$ and $l$. Then we take the average over the whole training dataset for the final gradient, which is used to update the variational parameters $\boldsymbol{\theta}$.
5) Repeat 2), 3), 4) for $s_0$ rounds. 
The pseudocode for our algorithm is provided in Supplementary Note 1.

In Fig.~\ref{fig:channnel_2^3}, we randomly choose $30$ different initial parameters $\boldsymbol{\theta}$, and then train DQNN$_1$ to learn the same target quantum channel. We observe that DQNN$_1$ converges quickly during the training process, with the highest fidelity above $96\%$. Compared with the numerical simulation results (see Supplementary Note 2), the deviation of the final converged fidelities is due to experimental imperfections, including qubit decoherence and residual $ZZ$ interactions between qubits~\cite{Zhao2020High,Ku2020Suppression,Kandala2021Demonstration}. 
In the upper left inset of Fig.~\ref{fig:channnel_2^3}, we show the distribution for all the converged fidelities from these $30$ repeated experiments. We expect that the distribution will concentrate to a higher fidelity for improved performance of the quantum processor. 

To evaluate the performance of DQNN$_1$, we choose one training process from the $30$ experiments, and refer the DQNN$_1$ with parameters corresponding to the ending (starting) epoch of the training curve as the trained (untrained) DQNN$_1$. We generate other $100$ different input quantum states and experimentally measure their corresponding output states produced by the trained (untrained) DQNN$_1$.
We test the fidelity between output states given by the target channel and the trained (untrained) DQNN$_1$.
As shown in the lower left inset of Fig.~\ref{fig:channnel_2^3}, for the trained DQNN$_1$, $43\%$ of the fidelities exceed $0.95$ (green curve) and $95\%$ of the fidelities are higher than $0.9$, which separate away from the distribution of the results of the untrained DQNN$_1$ (purple curve). This contrast illustrates the effectiveness of the training process of DQNN$_1$.

Another application of DQNNs is learning the ground state energy of a given Hamiltonian $H$ by minimizing the energy estimate  $\operatorname{tr}\left( \rho^{\text{out}} H \right)$ for the output state of the DQNN. 
Here we aim to learn the ground state energy of the molecular hydrogen Hamiltonian~\cite{OMalley2016Scalable}.
By exploiting the Bravyi-Kitaev transformation and certain symmetry, the Hamiltonian of molecular hydrogen can be reduced to the effective Hamiltonian acting on two qubits: $\hat{H}_{\mathrm{BK}} = g_{0} \mathbf{I}+g_{1} Z_{0}+g_{2} Z_{1}+g_{3} Z_{0} Z_{1}+g_{4} Y_{0} Y_{1}+g_{5} X_{0} X_{1} $, where $X_i, Y_i, Z_i$ are Pauli operators on the $i$-th qubit, and coefficients $g_j$ ($j=0,\cdots,5$) depend on the fixed bond length of molecular hydrogen. We consider the bond length $0.075$~nm in this work and the corresponding coefficients $g_i$ can be found in Ref.~\cite{OMalley2016Scalable}.

We use DQNN$_1$ again as the variational ansatz to learn the ground state of molecular hydrogen with the following procedure, similar to the previous one of learning a quantum channel:
1) Initialization: we prepare the input layer to the fiducial product state $|00\rangle$, and randomly generate initial gate parameters $\boldsymbol{\theta}$ for DQNN$_1$.
2) In the forward process (implemented on the quantum processor), we apply forward channels $\mathcal{E}^{1}$ and $\mathcal{E}^{\text {out}}$ in succession, and extract quantum states of the hidden layer ($\rho^{1}$) and the output layer ($\rho^{\text{out}}$) by quantum state tomography.
3) In the backward process (implemented on a classical computer), we initialize the quantum state of the output layer to $\sigma^{\text{out}}$,
and then obtain $\sigma^{1}$ and  $\sigma^{0}$ after successively applying backward channels $\mathcal{F}^{\text{out}}$ and $\mathcal{F}^{1}$ on $\sigma^{\text{out}}$.
4) Based on {$\{\left(\rho^{l-1},\sigma^{l}\right)\}$}, we calculate the gradient of the energy estimate with respect to all the variational parameters in the adjacent layers $l-1$ and $l$, and then update all gate parameters in DQNN$_1$.
5) Repeat 2), 3), 4) for $s_0$ rounds.
The pseudocode for our algorithm is provided in Supplementary Note 1.
\begin{figure*}[t]
\includegraphics{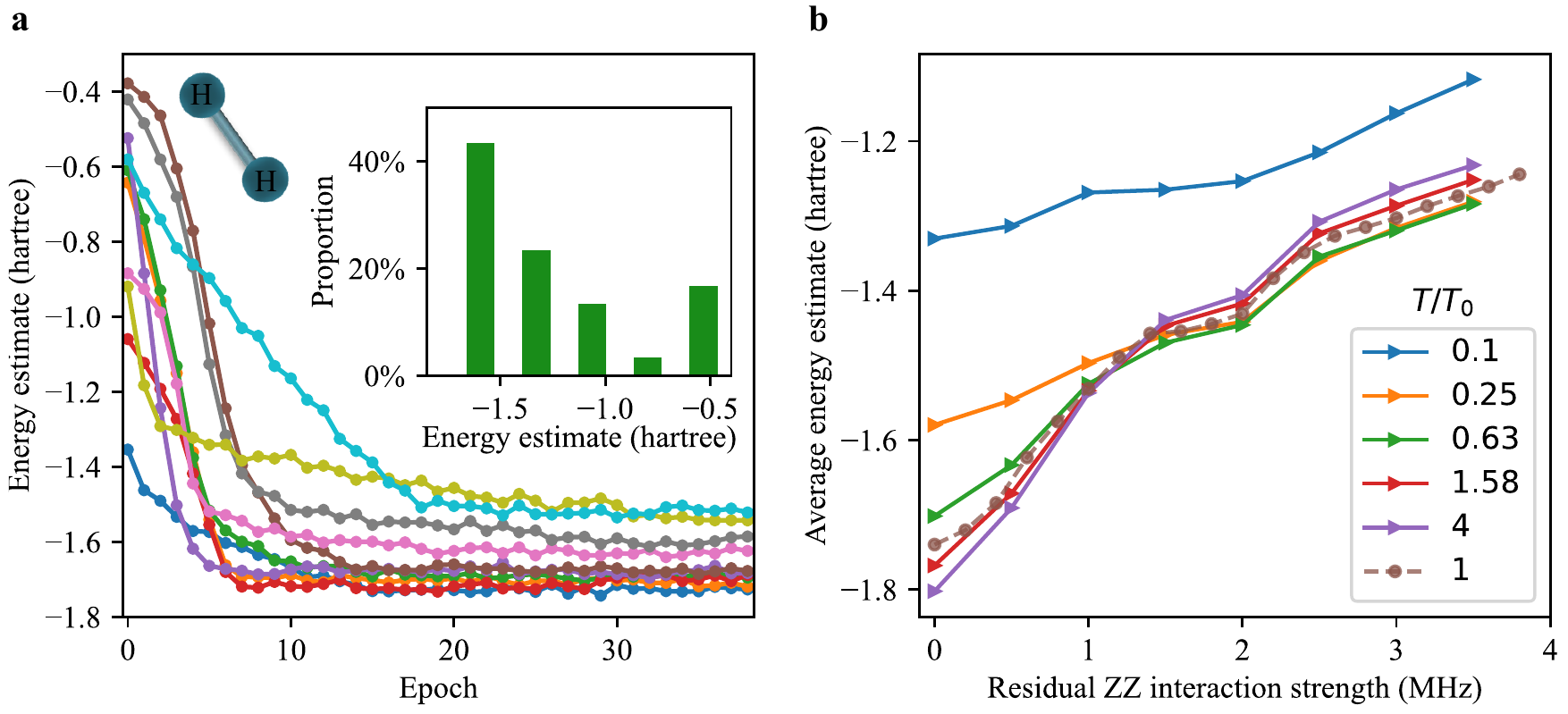}
\caption{
\textbf{Experimental and numerical results for learning the ground state energy of molecular hydrogen.} 
(a), Experimental energy estimate at each epoch during the learning process for different initial parameters.
The inset displays the distribution of converged energy estimates of $30$ different initial parameters.
(b), Numerical results for the mean energy estimates with different coherence times and residual $ZZ$ interaction strengths between qubits. Specifically, we adjust both the energy relaxation time and the dephasing time with the same ratio ($T/T_0$), where $T$ and $T_0$ are the coherence times in the simulation and the experiment respectively, and vary the residual $ZZ$ interaction strengths.
}
\label{fig:H2}
\end{figure*}

\begin{figure}[t]
\centering
\includegraphics{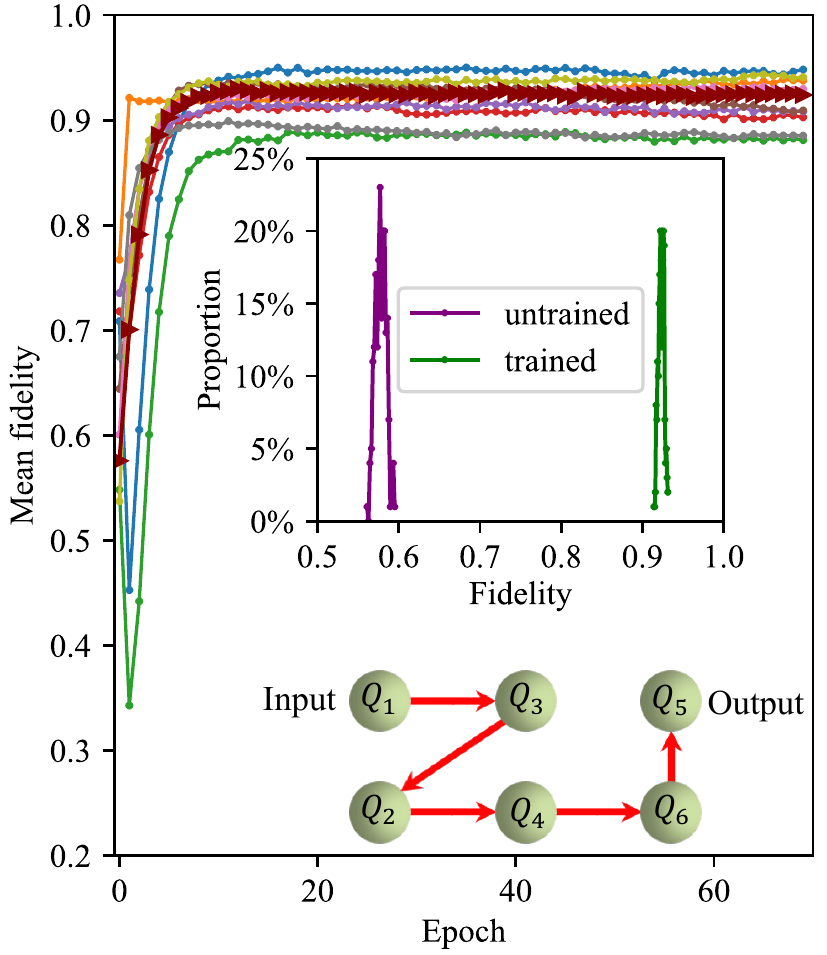}
\caption{
\textbf{Experimental results for learning a one-qubit quantum channel.} 
The mean fidelity of training the six-layer DQNN$_2$ is plotted as the function of training epochs for different initial parameters.
We randomly generate $100$ different single-qubit states, and evaluate the fidelities between their output states produced by DQNN$_2$ and their desired output states given by the target quantum channel.
The upper inset displays the distribution in two cases: a well-trained DQNN$_2$ (green) and an untrained (purple) DQNN$_2$, both are defined with the learning curve marked in triangles. The lower inset is a schematic illustration of DQNN$_2$, where we apply the perceptrons in the order indicated by the direction of the arrows.
}
\label{fig:channel_1^6}
\end{figure}

We train DQNN$_1$ with $30$ different initial parameters and show our experimental results in Fig.~\ref{fig:H2}(a).
We observe that DQNN$_1$ converges within $20$ epochs. The lowest ansatz energy estimate reaches below $-1.727$ (hartree) in the learning process,  with an accuracy up to $93.3\%$ compared to the theoretical value of the ground state energy $-1.851$ (hartree). 
This shows the good performance of DQNN$_1$  and the accuracy of our experimental system control. 
The inset of Fig.~\ref{fig:H2}(a) shows the distribution of all the converged energy from these $30$ repeated experiments with different initial parameters, 
six of which have an accuracy above $90\%$.   

To numerically investigate the effects of experimental imperfections on training DQNNs, we
consider two possible sources of errors: decoherence of qubits and residual $ZZ$ interactions between qubits. Taking into consideration these errors, we numerically train DQNN$_1$ with $30$ different initial parameters.
We find that for four of these initial parameters DQNN$_1$ converges to local minima instead of the global minimum, which is also observed in the experiment as shown in the inset of Fig.~\ref{fig:H2}(a).
Excluding these abnormal instances with local minima, we plot the average energy estimate as a function of the strength of the residual $ZZ$ interaction with different coherence times in Fig.~\ref{fig:H2}(b). We find that the increase of the coherence time around the experimental value has a minor effect on learning the ground state energy, while the reduction of the residual $ZZ$ interactions provides larger improvements of the ground state energy estimation.
These experimental imperfections can be suppressed after introducing advanced technologies in the design and fabrication of better superconducting quantum circuits, such as tunable couplers~\cite{Li2020Tunable,Collodo2020Implementation,Sung2021Realization} and tantalum based qubits~\cite{Place2021New,Wang2022Towards}.
 
To further illustrate the efficiency of the quantum backpropagation algorithm, we construct another DQNN  with four hidden layers (denoted as DQNN$_2$) by rearranging our six-qubit quantum processor into a six-layer structure, with one qubit respectively in each layer. We focus on the task of learning a one-qubit quantum channel. We choose $|0\rangle,|1\rangle,|-\rangle$ as our input states and compare the measured output states of DQNN$_2$ with the desired ones from the target single-qubit quantum channel.
The general training procedure is similar as in training DQNN$_1$ discussed above.
Our experimental results are summarized in Fig.~\ref{fig:channel_1^6}, which shows the learning curves for $10$ different initial parameters. We find that DQNN$_2$ can learn the target quantum channel with a mean fidelity up to $94.8\%$. We notice that the variance among the converged mean fidelity in DQNN$_2$ is smaller than that for DQNN$_1$, which may be attributed to the smaller total circuit depth and thus less error accumulation due to experimental imperfections. 
To study the learning performance, we choose one of these learning curves (marked in triangles), and refer DQNN$_2$ with parameters corresponding to the ending (starting) epoch of the learning curve as the trained (untrained) DQNN$_2$.  We then use other $100$ different input quantum states to test the trained and untrained DQNN$_2$ by measuring the fidelities between the experimental output states and the corresponding desired ones given by the target quantum channel. As shown in the upper inset of Fig.~\ref{fig:channel_1^6}, the fidelity distribution concentrates around $0.92$ for the trained DQNN$_2$, which stands in stark contrast to that of the untrained DQNN$_2$ and thus indicates a good performance after training. 

In summary, we have demonstrated the training of deep quantum neural networks on a six-qubit programmable superconducting quantum processor.
We experimentally exhibit its intriguing ability to learn quantum channels and learn the ground state energy of a given Hamiltonian. 
The quantum backpropagation algorithm demonstrated in our experiments can be directly applied to DQNNs with extended widths and depths.
This approach significantly reduces the requirements for the coherence time of superconducting qubits, regardless of how many hidden layers DQNNs include.
With further improvements in experimental conditions, the quantum perceptrons in our DQNNs can be constructed with deeper circuits to improve the expressive capacity, which allows DQNNs to tackle more challenging tasks in the future.

\vspace{.3cm}

\noindent\textbf{{Methods}}

\noindent\textbf{Framework.} 
We consider a deep quantum neural network (DQNN) that includes $L$ hidden layers with a total number $m_l$ of qubits in layer $l$. The qubits in two adjacent layers are connected with quantum perceptrons and each perceptron consists of two single-qubit rotation gates $R_x(\theta_1)$ and $R_x(\theta_2)$ along the $x$ axis with variational angles $\theta_1$ and $\theta_2$, respectively, followed by a fixed two-qubit controlled-phase gate.
The unitary of the quantum perceptron that acts on the $i$-th qubit at layer $l-1$ and the $j$-th qubit at layer $l$ in the DQNN is written as $U_{(i,j)}^{l}(\theta_{(i,j),1}^{l}, \theta_{(i,j),2}^{l})$.
Then the unitary product of all quantum perceptrons acting on the qubits in layers $l-1$ and $l$ is denoted as $U^{l}=\prod_{j=m_{l}}^{1} \prod_{i=m_{l-1}}^{1} U_{(i,j)}^{l}$.
The DQNN acts on the input state $\rho^{\text{in}}$ and produces the output state $\rho^{\text{out}}$ according to 
\begin{equation}
\rho^{\text {out}} \equiv \operatorname{tr}_{\text {in,hid}}\left(\mathcal{U}\left(\rho^{\text {in}} \otimes|0 \cdots 0\rangle_{\text {hid,out}}\langle 0 \cdots 0|\right) \mathcal{U}^{\dagger}\right),
\end{equation}
where $\mathcal{U} \equiv U^{\text{out}} U^{L} U^{L-1} \ldots U^{1}$ is the unitary of the DQNN,
and all qubits in the hidden layers and the output layer are initialized to a fiducial product state $|0 \cdots 0\rangle$. The characteristic of the layer-by-layer architecture enables $\rho^{\text{out}}$ to be expressed as a series of maps on $\rho^{\text{in}}$:
\begin{equation}
\rho^{\text{out}}=\mathcal{E}^{\text {out}}\left(\mathcal{E}^{L}\left(\ldots \mathcal{E}^{2}\left(\mathcal{E}^{1}\left(\rho^{\text {in}}\right)\right) \ldots\right)\right) ,
\end{equation}
where $\mathcal{E}^{l}\left(\rho^{l-1}\right) \equiv \operatorname{tr}_{l-1}\left( U^{l}\left(\rho^{l-1} \otimes|0 \cdots 0\rangle_{l}\langle 0 \cdots 0|\right) U^{l^{\dagger}} \right)$ is the forward quantum channel.

In Supplementary Information, we prove that for the two machine learning tasks in our work, the derivative of the mean fidelity or the energy estimate with respect to $\theta_{(i,j),k}^{l}$ can be calculated with the information of layers $l-1$ and $l$, which can be written as  $G(\boldsymbol{\theta^{l}}, \rho^{l-1}, \sigma^{l}) $ with $\boldsymbol{\theta^{l}}$ incorporating all parameters in layers $l-1$ and $l$.
We note that
$\rho^{l-1} = \mathcal{E}^{l-1}\left(\ldots \mathcal{E}^{2}\left(\mathcal{E}^{1}\left(\rho^{\text {in}}\right)\right) \ldots\right)$ refers to the quantum state in layer $l-1$ in the forward process, and $\sigma^{l} = \mathcal{F}^{l+1}\left(\ldots \mathcal{F}^{\text{out}}\left( \cdots \right) \ldots\right) $ represents the backward term in layer $l$ with $\mathcal{F}^{l}$ being the adjoint channel of  $\mathcal{E}^{l}$.
\\ \hspace*{\fill} \\
\noindent\textbf{Generating random input quantum states.}
To evaluate the learning performance in the task of learning a target quantum channel, we need to generate many different input quantum states and test the fidelity between their output states produced by DQNN$_1$ and their desired output states given by the target quantum channel.

For the task of learning a two-qubit quantum channel, we generate these input quantum states by separately applying single-qubit rotation gates $R_{a_1}(\Omega_1)\otimes{}R_{a_2}(\Omega_2)$ on the two qubits initialized in $\left|00\right\rangle$. Here each rotation gate has a random rotation axis $a_i$ in the $x$-$y$ plane and a random rotation angle $\Omega_i$.

For the task of learning a one-qubit quantum channel, we generate the input quantum states by applying single-qubit rotation gates $R_{b}(\Phi)$ on the input qubit initialized in $\left|0\right\rangle$ with a random rotation axis $b$ in the $x$-$y$ plane and a random rotation angle $\Phi$.

\vspace{.3cm}
\noindent\textbf{{Data availability}}
The data for experimental results presented in the figures is provided in 
\url{https://github.com/luzd19/Deep-quantum-neural-networks_equipped-with-backpropagation}

\vspace{.3cm}
\noindent \textbf{Code Availability}
The codes for numerical simulations and the numerical results are available at 
\url{https://github.com/luzd19/Deep-quantum-neural-networks_equipped-with-backpropagation}

\vspace{.3cm}
\noindent \textbf{Acknowledgement}
We thank Wenjie Jiang for helpful discussions. 
We acknowledge the support by the National Natural Science Foundation of China (Grants No.~92165209, No.~11925404, No.~11874235, No.~11874342, No.~11922411, No.~12061131011, No.~T2225008, No.~12075128), the National Key Research and Development Program of China (Grants No.~2017YFA0304303), Key-Area Research and Development Program of Guangdong Province (Grant No.~2020B0303030001), Anhui Initiative in Quantum Information Technologies (AHY130200), China Postdoctoral Science Foundation (BX2021167), and Grant No.~2019GQG1024 from the Institute for Guo Qiang, Tsinghua University. D.-L. D. also acknowledges additional support from the Shanghai Qi Zhi Institute.

\vspace{.3cm}
\noindent \textbf{Author contributions} 
X.P. carried out the experiments and analyzed the data with the assistance of Z.H. and Y.X.; L.S. directed the experiments; 
Z.L. formalized the theoretical framework and performed the numerical simulations under the supervision of D.-L.D.;
W.L. and C.-L.Z. provided theoretical support;
W.C. fabricated the parametric amplifier;
W.W. and X.P. designed the devices; 
X.P. fabricated the devices with the assistance of W.W., H.W., and Y.-P.S.;
Z.H., W.C., and X.L. provided further experimental support;
X.P., Z.L., W.L., C.-L.Z., D.-L.D., and L.S. wrote the manuscript with feedback from all authors.

\vspace{.3cm}
\noindent \textbf{Competing interests} 
All authors declare no competing interests.

\bibliographystyle{Zou}
\bibliography{DQNN_lzd, DQNN_pxx}


\clearpage
\newpage 
\onecolumngrid
\setcounter{section}{0}
\setcounter{equation}{0}
\setcounter{figure}{0}
\setcounter{table}{0}
\setcounter{page}{1}
\makeatletter
\renewcommand\thefigure{S\arabic{figure}}
\renewcommand\thetable{S\arabic{table}}
\renewcommand\theequation{S\arabic{equation}}
\renewcommand{\figurename}{Supplementary Figure}

\begin{center} 
	{\large \bf Supplementary Information: Deep quantum neural networks equipped with backpropagation on a superconducting processor}
\end{center}

\maketitle

\section*{SUPPLEMENTARY NOTE 1: Theoretical details for deep quantum neural networks}

In classical machine learning, deep neural networks are characterized by the ability to extract high-level features from data. With the rapid development in quantum machine learning~\cite{Biamonte2017Quantum, Dunjko2018Machine, DasSarma2019Machine, Cerezo2022Challenges, Dawid2022Modern}, we expect a quantum generalization of a deep neural network architecture to bring promising insights.
Recently, a deep quantum neural network (DQNN) and a quantum analog of the backpropagation algorithm have been proposed~\cite{Beer2020Training}. In this ansatz, the quantum analog of a perceptron is a unitary operation which acts on qubits in two adjacent layers. During the training process, the unitary operator of a quantum perceptron is updated by multiplying the corresponding updating matrix.

In this paper, we experimentally demonstrate the training of parameterized DQNNs with a superconducting quantum processor. 
Equipped with the backpropagation algorithm, we can efficiently calculate the gradients during the training process. Our scheme is feasible for the experimental implementation in the noisy intermediate scale quantum era.
In this section, we will introduce the basic structures, optimization strategies, and training procedures for DQNNs.

\subsection*{Basic structures}
As mentioned in the main text, our DQNNs have layer-by-layer structures, and
qubits in two adjacent layers are connected with the quantum perceptrons. 
In our ansatz, the quantum perceptrons are engineered as parameterized quantum circuits.
For simplicity in this paper, we consider that each quantum perceptron acts on only two qubits in two adjacent layers. 
The circuit structure of a quantum perceptron is composed of two single-qubit rotation gates $R_x(\theta_1)$ and $R_x(\theta_2)$ with $\theta_1$ and $\theta_2$ as the variational parameters, followed by a fixed two-qubit controlled-phase gate, which is shown in Fig.~1(b) in the main text.
A sequential combination of the quantum perceptrons constitutes the layer-by-layer transition mapping between adjacent layers.
In this way, the DQNN maps the information layerwise from the input layer to the output layer through hidden layers.

Now we consider a DQNN including $L$ hidden layers. The total number of qubits in layer $l$ is denoted as $m_l$.
The unitary of a quantum perceptron which acts on the $i$-th qubit at layer $l-1$ and the $j$-th qubit at layer $l$ is written as $U_{(i,j)}^{l}(\theta_{(i,j),1}^{l}, \theta_{(i,j),2}^{l})$, where $\theta_{(i,j),k}^{l}$ ($k=1,2$)  denote the variational parameters of the two $R_x$ gates in the quantum perceptron $U_{(i,j)}^{l}$.
The unitary product of all quantum perceptrons acting on the qubits in layers $l-1$ and $l$ is denoted as:
\begin{equation*}
U^{l}=\prod_{j={m_l}}^{1} \prod_{i=m_{l-1}}^{1} U_{(i,j)}^{l}  .
\end{equation*}
We note that qubits in layer $l$ are initialized to a fiducial product state $|0 \cdots 0\rangle$, and then the quantum state $\rho^l$ of the qubits in layer $l$ can be written as the layer-by-layer transition mapping on $\rho^{l-1}$:
\begin{equation}
\rho^l = \mathcal{E}^{l}\left(\rho^{l-1}\right) \equiv \operatorname{tr}_{l-1}\left( U^{l}\left(\rho^{l-1} \otimes|0 \cdots 0\rangle_{l}\langle 0 \cdots 0|\right) U^{l^{\dagger}} \right) .
\end{equation}
In this way, the output state $\rho^{\text{out}}$ can be expressed as a series of maps on $\rho^{\text{in}}$:
\begin{equation}
\rho^{\text{out}}=\mathcal{E}^{\text {out}}\left(\mathcal{E}^{L}\left(\ldots \mathcal{E}^{2}\left(\mathcal{E}^{1}\left(\rho^{\text {in}}\right)\right) \ldots\right)\right) .
\end{equation}

\subsection*{Optimization strategies}

With the basic structures discussed above, now we can specify the learning tasks.
In this paper, we consider two machine learning tasks. 
The first task is learning a target quantum channel. We expect the output state given by the DQNN to be as close as possible to the output state given by the target quantum channel for each input state.
We aim to maximize the mean fidelity between output states given by the DQNN ($\rho_x^{\text{out}}$) and the target quantum channel ($\tau_x^{\text {out}}$) averaged over $N$ training data:
\begin{equation*}
F = \frac{1}{N}\sum_{x=1}^{N} F_x(\rho^{\text {out}}_x, \tau^{\text {out}}_x) = \frac{1}{N}\sum_{x=1}^{N}\left[\operatorname{tr} \sqrt{\sqrt{\tau_{x}^{\text {out}}} \rho_{x}^{\text{out }} \sqrt{\tau_{x}^{\text {out}}}}\right] .
\end{equation*}
The second task is learning the ground state of a Hamiltonian $H$.  We aim to minimize the energy estimate $\bar{E}$ of this Hamiltonian computed with the DQNN output state $\rho^{\text{out}}$:
\begin{equation*}
\bar{E} = \operatorname{tr}\left( \rho^{\text{out}} H \right).
\end{equation*}
To maximize the mean fidelity or minimize the energy estimate, we adapt the gradient descent method.
In the main text, we mention that our DQNNs with the layer-by-layer architecture allow the quantum backpropagation algorithm. Via this algorithm, one only requires the information from two adjacent layers to calculate the gradients with respect to all gate parameters at these two layers. 
In other words, the derivative of the mean fidelity or the energy estimate with respect to $\theta_{(i,j),k}^{l}$ can be written as  $G(\boldsymbol{\theta^{l}}, \rho^{l-1}, \sigma^{l}) $, where $\boldsymbol{\theta^{l}}$ incorporates all gate parameters in layers $l-1$ and $l$.

Here, we derive the formula for $G(\boldsymbol{\theta^{l}}, \rho^{l-1}, \sigma^{l}) $.
We first consider a function $f$ with the form $f(\rho^{\text{out}},X)=\operatorname{tr}( X \rho^{\text{out}})$, where $X$ is a Hermitian matrix related to specific tasks.
The derivative of $f$ with respect to $\theta_{(i,j),k}^{l}$ can be expressed as:
\begin{equation}
\begin{aligned}
\frac{\partial f(\rho^{\text{out}}, X)}{\partial \theta_{(i,j),k}^{l}} 
&= G(\boldsymbol{\theta^{l}}, \rho^{l-1}, \sigma^{l}) \\
&= \operatorname{tr}  \bigg(  \frac{\partial U^{l}}{\partial \theta_{(i,j),k}^{l}}  \Big( \rho^{l-1} \otimes |0\rangle_{l}\langle 0| \Big) U^{l^{\dagger}}     \Big( \mathbb{I}_{l-1}\otimes \sigma^{l}  \Big)   \bigg) + \mathrm{h.c.} ,
\end{aligned}
\end{equation}
where $\mathrm{h.c.}$ stands for the Hermitian conjugate of the preceding terms. We show the proof as follows: \\ \\
\textbf{Proof.}
\begin{align*}
\frac{\partial f(\rho^{\text{out}}, X)}{\partial \theta_{(i,j),k}^{l}} 
&= G(\boldsymbol{\theta^{l}}, \rho^{l-1}, \sigma^{l}) = \operatorname{tr}\bigg( \frac{\partial \rho^{\text{out}}}{\partial \theta_{(i,j),k}^{l}} X \bigg) \\
&= \operatorname{tr} \bigg( \Big(\operatorname{tr}_{\text{in,hidden}}\Big(U^{\text{out}}  \ldots U^{l+1} \frac{\partial U^{l}}{\partial \theta_{(i,j),k}^{l}} U^{l-1} \ldots U^{1} \rho^{\text{in}} \otimes \left|0 \cdots 0\right \rangle_{\text{hid},\text{out}}\left \langle 0 \cdots 0 \right |  \mathcal{U} ^{\dagger}\Big) + \mathrm{h.c.} \Big) X \bigg) \\
&= \operatorname{tr} \bigg( \Big( U^{\text{out}}  \ldots U^{l+1} \frac{\partial U^{l}}{\partial \theta_{(i,j),k}^{l}} U^{l-1} \ldots U^{1} \rho^{\text{in}} \otimes \left |0 \cdots 0\right \rangle_{\text{hid},\text{out}}\left \langle 0 \cdots 0\right |  \mathcal{U} ^{\dagger} \Big) \cdot \Big(  \mathbb{I}_{\text{in,hidden}} \otimes X \Big)  \bigg) + \mathrm{h.c.} \\
&= \operatorname{tr} \bigg(  \frac{\partial U^{l}}{\partial \theta_{(i,j),k}^{l}} U^{(l-1)} \ldots U^{1} \rho^{\text{in}} \otimes |0 \cdots 0\rangle_{\text{hid},\text{out}}\langle 0 \cdots 0| U^{1^{\dagger}} \ldots U^{(l-1)^{\dagger}}  \\
&\quad  U^{(l)^{\dagger}} U^{(l+1)^{\dagger}} \ldots U^{\text{out}^{\dagger}}  \Big(  \mathbb{I}_{\text{in,hidden}} \otimes X  \Big)  U^{\text{out}} \ldots U^{(l+1)}   \bigg ) + \mathrm{h.c.} \\
&=  \operatorname{tr} \bigg( \frac{\partial U^{l}}{\partial \theta_{(i,j),k}^{l}} \Big(  \operatorname{tr}_{l,...\text{out}} \left( T_1 \right) \otimes |0 \cdots 0\rangle_{l,...,\text{out}}\langle 0 \cdots 0|    \Big) U^{l^{\dagger}} \Big( \mathbb{I}_{0,...,l-1} \otimes \operatorname{tr} _{0,...l-1} \left(  T_2            \right)           \Big)         \bigg) + \mathrm{h.c.}  \\
&=  \operatorname{tr} \Bigg( \bigg(  \Big( \frac{\partial U^{l}}{\partial \theta_{(i,j),k}^{l}}  \big( \operatorname{tr}_{l,...\text{out}} \left( T_1 \right) \otimes |0\rangle_{l}\langle 0| \big) U^{l^{\dagger}} \Big) \otimes \mathbb{I}_{l+1,...\text{out}} \bigg)  \\
&\quad  \bigg( \mathbb{I}_{0,...l}\otimes |0\rangle_{l+1,...\text{out}}\langle 0|  \bigg)       \bigg(  \mathbb{I}_{0,...l-1} \otimes \operatorname{tr} _{0,...l-1} \left(  T_2                    \right)    \bigg)   \Bigg) +  \mathrm{h.c.} \\
&=  \operatorname{tr} \Bigg( \bigg(  \Big( \frac{\partial U^{l}}{\partial \theta_{(i,j),k}^{l}}  \big( \rho^{l-1} \otimes |0\rangle_{l}\langle 0| \big)  U^{l^{\dagger}} \Big)  \otimes \mathbb{I}_{l+1,...\text{out}} \bigg) \bigg( \mathbb{I}_{l-1,l}\otimes |0\rangle_{l+1,...\text{out}}\langle 0|  \bigg)  \bigg(  \mathbb{I}_{l-1} \otimes \operatorname{tr} _{0,...l-1} \left(  T_2 \right)    \bigg) \Bigg) +  \mathrm{h.c.} \\
&=  \operatorname{tr}  \bigg(  \frac{\partial U^{l}}{\partial \theta_{(i,j),k}^{l}}  \Big( \rho^{l-1} \otimes |0\rangle_{l}\langle 0| \Big) U^{l^{\dagger}}    \Big( \mathbb{I}_{l-1}\otimes \sigma^{l} \Big)     \bigg) +  \mathrm{h.c.} ,
\end{align*}
where 
$\mathcal{U} \equiv U^{\text{out}} U^{L} U^{L-1} \ldots U^{1}$. We use the shorthands $T_1 = U^{(l-1)} \ldots U^{1} \rho^{\text{in}} \otimes |0 \cdots 0\rangle_{\text{hid},\text{out}}\langle 0 \cdots 0| U^{1^{\dagger}} \ldots U^{(l-1)^{\dagger}} $, and $T_2 = 1/(2^{\sum_{i=m_{0}}^{m_{l-1}}}) U^{(l+1)^{\dagger}} \ldots U^{\text{out}^{\dagger}}  \left(  \mathbb{I}_{\text{in,hidden}}  \otimes X  \right)   U^{\text{out}} \ldots U^{(l+1)} $. 
We define $\rho^{l-1} = \operatorname{tr}_{1,...,l-2,l,...,\text{out}} \left( T_1 \right) $ as the quantum states of the qubits in layer $l-1$ in the forward process, and $\sigma^{l} = \operatorname{tr}_{l+1,...,\text{out}}\left( \left( \mathbb{I}_{l} \otimes |0 \cdots 0\rangle_{l+1,...,\text{out}}\langle 0 \cdots 0|  \right)  \cdot  \operatorname{tr} _{1,...,l-1} \left(  T_2 \right)  \right)  $ as the backward term in layer $l$. From this formula we obtain the recursive relation between $\sigma^{l-1}$ and $\sigma^{l}$:     
\begin{equation} 
\sigma^{l-1} = \mathcal{F}^{l}(\sigma^{l})  =  \operatorname{tr}_{l} \left(  \left( \mathbb{I}_{l-1} \otimes |0\rangle_{l}\langle 0|  \right)  U^{l^{\dagger}} \left(   \mathbb{I}_{l-1}\otimes \sigma^{l}   \right)  U^{l}  \right) ,
\end{equation}  
where $\mathcal{F}^{l}$ is the adjoint channel of $\mathcal{E}^{l}$, and $\sigma^{\text{out}} = X $.
From this recursive relation, we can obtain the backward terms layerwise from the output layer to the input layer in the backward process.

Specially, if the gate with parameter $\theta_{(i,j),k}^{l}$ in the DQNN is of the form $e^{-\frac{i}{2}\theta_{(i,j),k}^{l}P_n}$ with $P_n$ belonging to the Pauli group, we can utilize the ``parameter shift rule'' to calculate the gradient of $f$:
\begin{equation}
\frac{\partial f(\rho, \sigma)}{\partial \theta_{(i,j),k}^{l}} = G(\boldsymbol{\theta^{l}}, \rho^{l-1}, \sigma^{l})
=  \frac{1}{2}(h_{+} - h_{-})   , 
\end{equation}
where $ h_{\pm} =  \operatorname{tr}  \left(  U^{l}_{\pm} \left( \rho^{l-1} \otimes |0\rangle_{l}\langle 0| \right) U^{l^{\dagger}}_{\pm}    \left( \mathbb{I}_{l-1}\otimes \sigma^{l} \right)   \right) $,
and $U^{l}_{\pm}$ denotes the unitary that replaces the parameter $\theta_{(i,j),k}^{l}$ in $U^{l}$ with $\theta_{(i,j),k}^{l} \pm \frac{\pi}{2}$. We show the proof as follows: \\ \\
\textbf{Proof.}
\begin{align*}
2 \cdot \frac{\partial f(\rho^{\text{out}}, X)}{\partial \theta_{(i,j),k}^{l}} 
&= 2 \cdot  G(\boldsymbol{\theta^{l}}, \rho^{l-1}, \sigma^{l}) = 2 \cdot \operatorname{tr}\bigg(\frac{\partial \rho^{\text{out}}}{\partial \theta_{(i,j),k}^{l}} X \bigg) \\
&= \operatorname{tr} \Bigg( \bigg(\operatorname{tr}_{\text{in,hidden}}\Big(U^{\text{out}}  \ldots U^{l+1}   U^{l}_{+}     U^{l-1} \ldots U^{1} \rho^{\text{in}} \otimes |0 \cdots 0\rangle_{\text{hid},\text{out}}\langle 0 \cdots 0|  U^{1^{\dagger}} \cdots U^{l^{\dagger}}_{+} \cdots U^{\text{out}^{\dagger}} \Big) \bigg) X \Bigg) \\
&-  \operatorname{tr} \Bigg( \bigg(\operatorname{tr}_{\text{in,hidden}}\Big(U^{\text{out}}  \ldots U^{l+1}   U^{l}_{-}     U^{l-1} \ldots U^{1} \rho^{\text{in}} \otimes |0 \cdots 0\rangle_{\text{hid},\text{out}}\langle 0 \cdots 0|  U^{1^{\dagger}} \cdots U^{l^{\dagger}}_{-} \cdots U^{\text{out}^{\dagger}} \Big) \bigg) X \Bigg) .
\end{align*}
Now we prove the first term equals to $h_{+}$. In the same way, we can prove the second term equals to $h_{-}$.
The first term can be written as:
\begin{align*}
&\quad \operatorname{tr} \Bigg( \bigg(\operatorname{tr}_{\text{in,hidden}}\Big(U^{\text{out}}  \ldots U^{l+1}   U^{l}_{+}     U^{l-1} \ldots U^{1} \rho^{\text{in}} \otimes |0 \cdots 0\rangle_{\text{hid},\text{out}}\langle 0 \cdots 0|  U^{1^{\dagger}} \cdots U^{l^{\dagger}}_{+} \cdots U^{\text{out}^{\dagger}} \Big) \bigg) X \Bigg) \\
&= \operatorname{tr} \Bigg(  \bigg( U^{\text{out}}  \ldots U^{l+1}   U^{l}_{+}     U^{l-1} \ldots U^{1} \rho^{\text{in}} \otimes |0 \cdots 0\rangle_{\text{hid},\text{out}}\langle 0 \cdots 0|  U^{1^{\dagger}} \cdots U^{l^{\dagger}}_{+} \cdots U^{\text{out}^{\dagger}}   \bigg)  \bigg(\mathbb{I}_{\text{in,hidden}} \otimes X \bigg)  \Bigg) \\
&= \operatorname{tr} \bigg(  U^{l}_{+} U^{(l-1)} \ldots U^{1} \rho^{\text{in}} \otimes |0 \cdots 0\rangle_{\text{hid},\text{out}}\langle 0 \cdots 0| U^{1^{\dagger}} \ldots U^{(l-1)^{\dagger}}  U^{l^{\dagger}}_{+} U^{(l+1)^{\dagger}} \ldots U^{\text{out}^{\dagger}} \\
&\qquad \Big(  \mathbb{I}_{\text{in,hidden}} \otimes X  \Big)  U^{\text{out}} \ldots U^{(l+1)}   \bigg)  \\
&=  \operatorname{tr} \bigg( U^{l}_{+}  \Big(  \operatorname{tr}_{l,...\text{out}} \left( T_1 \right) \otimes |0 \cdots 0\rangle_{l,...\text{out}}\langle 0 \cdots 0|    \Big) U^{l^{\dagger}}_{+} \Big(  \mathbb{I}_{0,...l-1} \otimes \operatorname{tr} _{0,...l-1} \left(  T_2   \right)    \Big)         \bigg)   \\
&=  \operatorname{tr} \Bigg( \bigg(  \Big( U^{l}_{+}  \left( \operatorname{tr}_{l,...\text{out}} \left( T_1 \right) \otimes |0\rangle_{l}\langle 0| \right) U^{l^{\dagger}}_{+} \Big)  \otimes \mathbb{I}_{l+1,...\text{out}} \bigg) \bigg(\mathbb{I}_{0,...l}\otimes |0\rangle_{l+1,...\text{out}}\langle 0|  \bigg) \\
&\qquad \bigg(  \mathbb{I}_{0,...l-1} \otimes \operatorname{tr} _{1,...l-1} \left(  T_2    \right)    \bigg)  \Bigg)   \\
&= \operatorname{tr} \Bigg( \bigg(  \Big( U^{l}_{+}  \left( \rho_{l-1} \otimes |0\rangle_{l}\langle 0| \right)  U^{l^{\dagger}}_{+} \Big)  \otimes \mathbb{I}_{l+1,...\text{out}} \bigg) \bigg( \mathbb{I}_{l-1,l}\otimes |0\rangle_{l+1,...\text{out}}\langle 0|  \bigg)      \bigg(  \mathbb{I}_{l-1} \otimes \operatorname{tr} _{0,...,l-1} \left(  T_2 \right)    \bigg)  \Bigg)  \\
&=  \operatorname{tr}  \left(  U^{l}_{+} \left( \rho_{l-1} \otimes |0\rangle_{l}\langle 0| \right) U^{l^{\dagger}}_{+}    \left( \mathbb{I}_{l-1}\otimes \sigma_{l} \right)   \right)  \\
&=  h_{+}  .
\end{align*}

With the gradients of $f$ obtained above, we need to derive the gradients of the mean fidelity $F$ and the energy estimate $\bar{E}$ in the two tasks that are discussed in the main text. 
In the task of learning a target quantum channel, for each input state, we consider the derivative of $F_x$ with respect to $\theta_{(i,j),k}^{l}$. 
For convenience, we omit the superscript and subscript of $F_x$, $\rho^{\text {out}}_x$, and $\tau^{\text {out}}_x$, and use the shorthand $A=\tau^{1/2} \rho \tau^{1/2}$, $ B=\sqrt{A}$, then
\begin{equation*}
\frac{\partial F}{\partial \theta_{(i,j),k}^{l}}  = \operatorname{tr}\left(\frac{\partial \mathrm{B}}{\partial \theta_{(i,j),k}^{l}} \right).
\end{equation*}
Now, we further omit the superscript and subscript of $\theta_{(i,j),k}^{l}$.
\begin{equation*}
\begin{gathered}
\frac{\partial \mathrm{A}}{\partial \theta} = \frac{\partial\left(\mathrm{B}^{2}\right)}{\partial \theta} = B \cdot \frac{\partial \mathrm{B}}{\partial \theta} + \frac{\partial \mathrm{B}}{\partial \theta} \cdot B   \Rightarrow   \frac{\partial \mathrm{A}}{\partial \theta} \cdot B^{-1} = B \cdot \frac{\partial \mathrm{B}}{\partial \theta} B^{-1} + \frac{\partial \mathrm{B}}{\partial \theta} ,
\end{gathered}
\end{equation*}
hence,
\begin{equation*}
\operatorname{tr}\left(\frac{\partial \mathrm{A}}{\partial \theta} \cdot B^{-1}\right) = \operatorname{tr}\left(B \cdot \frac{\partial \mathrm{B}}{\partial \theta} B^{-1}\right) + \operatorname{tr}\left(\frac{\partial \mathrm{B}}{\partial \theta}\right) = \operatorname{tr}\left(\frac{\partial \mathrm{B}}{\partial \theta} B^{-1} B\right)+\operatorname{tr}\left(\frac{\partial \mathrm{B}}{\partial \theta}\right) = 2 \operatorname{tr}\left(\frac{\partial \mathrm{B}}{\partial \theta}\right) .
\end{equation*}
This yields 
\begin{equation*}
\begin{aligned}
\frac{\partial F}{\partial \theta} &= \frac{1}{2} \operatorname{tr}\left(\frac{\partial \mathrm{A}}{\partial \theta} \cdot B^{-1}\right) = \frac{1}{2} \operatorname{tr}\left(\tau^{1/2} \frac{\partial \rho}{\partial \theta} \tau^{1/2} \cdot B^{-1}\right) = \frac{1}{2} \operatorname{tr}\left(\frac{\partial \rho}{\partial \theta} \cdot \tau^{1/2}  B^{-1} \tau^{1/2}\right) ,
\end{aligned}
\end{equation*}
which has the same form as the derivative of  $\operatorname{tr}(\rho^{\text{out}}X) $ with $\tau^{1/2}  B^{-1} \tau^{1/2}$ analogous to $X$.
In the task of learning the ground state energy of a Hamiltonian $H$, the energy estimate $\operatorname{tr}\left( \rho^{\text{out}} H \right)$ has the same form as  $\operatorname{tr}\left( \rho^{\text{out}} X \right)$, where $H$ is analogous to $X$. 
So we can derive the gradients of the energy estimate $\bar{E}$ according to $G(\boldsymbol{\theta^{l}}, \rho^{l-1}, \sigma^{l})$.
With the gradients obtained, we can update the variational parameters in the DQNN by gradient descent methods.

\subsection*{Training procedures}
In this section, we give a detailed description of how our DQNNs are trained via the quantum backpropagation algorithm for different tasks.

For the task of learning a quantum channel, first we need to generate the training dataset. 
Here, we randomly choose parameters $\boldsymbol{\theta_t}$ in the DQNN to generate a specific target quantum channel that we aim to learn. Then we apply the target quantum channel on each input state to obtain the corresponding output state to constitute the training dataset $\{\left(\rho_{x}^{\text{in}},\tau_{x}^{\text{out}}\right)\}^{N}_{x=1}$ with $N$ being the size of the training dataset.
We assume the DQNN used in this task includes $L$ hidden layers with a total number $m_l$ of qubits in layer $l$.
Now we describe the general training procedure as follows:\\

1. Initialization: \\
\indent \indent Randomly choose initial gate parameters for all perceptrons in the DQNN, which is denoted as $\boldsymbol{\theta_I}$ .

2. Forward process: \\
\indent \indent For each training data $\{\left(\rho_{x}^{\text {in}},\tau_{x}^{\text {out }}\right)\}$, apply forward channels $\mathcal{E}^{1}$, $\mathcal{E}^{2}$,$\ldots$, $\mathcal{E}^{\text{out}}$ on $\rho_{x}^{\text{in}}$ to obtain $\rho^{1}_x, \rho^{2}_x, \ldots, \rho^{\text{out}}_x$ successively .

\begin{itemize}  
\item[]
\textbf{Forward channel $\mathcal{E}^{l}$} : 
According to the main text, the forward channel $\mathcal{E}^{l}$ applies on qubits in layer $l-1$ of the quantum state $\rho^{l-1}$, and produces $\rho^{l}$ in layer $l$ according to 
$\rho^{l} = \mathcal{E}^{l}\left(\rho^{l-1}\right) \equiv \operatorname{tr}_{l-1}\left( U^{l}\left(\rho^{l-1} \otimes|0 \cdots 0\rangle_{l}\langle 0 \cdots 0|\right) U^{l^{\dagger}} \right)$.
In our experiment, we prepare $m_l$ qubits in layer $l$ to the fiducial product state $\left|0 \cdots 0 \right\rangle$ at first. Then we apply all quantum perceptrons acting on qubits in layers $l-1$ and $l$.
Finally, we carry out quantum state tomography to extract $\rho^{l}$.
\end{itemize}

3. Backward process: \\
\indent \indent For each training data $\{\left(\rho_{x}^{\text {in}},\tau_{x}^{\text {out}}\right)\}$, calculate $\sigma^{\text{out}} = (\tau_{x}^{\text{out}})^{1/2} ((\tau_{x}^{\text{out}})^{1/2} \rho_{x}^{\text{out}} (\tau_{x}^{\text{out}})^{1/2})^{-1/2} (\tau_{x}^{\text{out}})^{1/2}$, and then apply 
\indent \indent  backward channels $\mathcal{F}^{\text{out}}$, $\mathcal{F}^{L}$,$\ldots$, $\mathcal{F}^{1}$ on $\sigma^{\text{out}}$ to successively obtain $\sigma^{L}_x, \sigma^{L-1}_x, \ldots,\sigma^{0}_x $ .

\begin{itemize}  
\item[]
\textbf{Backward channel $\mathcal{F}^{l}$} :
The backward channel $\mathcal{F}^{l}$ applies on backward term $\sigma^{l}$ and produces $\sigma^{l-1}$ according to $\sigma^{l-1} = \mathcal{F}^{l}(\sigma^{l}) =  \operatorname{tr}_{l} \left(  \left( \rho_{l-1} \otimes |0\rangle_{l}\langle 0|  \right)                   U^{l^{\dagger}} \left( \mathbb{I}_{l-1}\otimes \sigma^{l} \right) U^{l} \right) $. 
In this paper, we carry out the backward channel on a classical computer due to the experimental challenges in preparing the quantum states for the backward terms $\sigma^{l}$. We expect an efficient proposal for the experimental implementation of the backward process, which is important and remains as a future work.
\end{itemize}

4. Evaluate the mean fidelity and the gradients: \\
\indent \indent Compute the mean fidelity:
\begin{equation*}
F = \frac{1}{N}\sum_{x=1}^{N} F_x(\rho^{\text {out}}_x, \tau^{\text {out}}_x) = \frac{1}{N}\sum_{x=1}^{N}\left[\operatorname{tr} \sqrt{\sqrt{\tau_{x}^{\text {out}}} \rho_{x}^{\text{out }} \sqrt{\tau_{x}^{\text {out}}}}\right] .
\end{equation*}
\indent \indent Calculate the gradient with respect to $\theta_{(i,j),k}^{l}$ for each training data:  $\frac{\partial F_x(\rho^{\text{out}}_x, \tau^{\text{out}}_x)}{\partial \theta_{(i,j),k}^{l}} = G(\boldsymbol{\theta^{l}}, \rho^{l-1}_x, \sigma^{l}_x)$, and then take the  
\indent \indent average over the whole training dataset: $ \frac{1}{N}\sum^N_{x=1} G(\boldsymbol{\theta^{l}}, \rho^{l-1}_x, \sigma^{l}_x) $.
Finally, update each $\theta_{(i,j),k}^{l}$ with the learning rate $\epsilon$ \indent \indent  according to 
\begin{equation*}
\theta_{(i,j),k}^{l} \rightarrow \theta_{(i,j),k}^{l} + \epsilon *  \frac{1}{N}\sum^N_{x=1} G(\boldsymbol{\theta^{l}}, \rho^{l-1}_x, \sigma^{l}_x).
\end{equation*}

5. Repeat $2$, $3$ and $4$ for $s_0$ steps.\\

We summarize the pseudocode in Algorithm~\ref{algorithm 1}. 
\begin{algorithm}[H]
\caption{Training the DQNN for learning quantum channels via the quantum backpropagation algorithm}
\label{algorithm 1}
\begin{algorithmic}
\REQUIRE The DQNN model with $L$ hidden layers, initial parameters $\boldsymbol{\theta_I}$, 
input quantum states $\{\left(\rho_{x}^{\text {in}}\right)\}^{N}_{x=1}$, 
iteration steps $s_0$ , learning rate $\epsilon$,
\ENSURE The trained DQNN
\STATE  \textbf{Generate the training dataset:} choose parameters $\boldsymbol{\theta_t}$ for the DQNN, which serves as the target quantum channel, and then apply it to each input state to obtain the corresponding output state, which constitute the training dataset $\{\left(\rho_{x}^{\text{in}},\tau_{x}^{\text{out}}\right)\}^{N}_{x=1}$ .
\FOR{ $s=1$ to $s_0$}
     \STATE \textbf{Forward:} for each training data $\{\left(\rho_{x}^{\text {in}},\tau_{x}^{\text {out }}\right)\}$, apply forward channels $\mathcal{E}^{1}$, $\mathcal{E}^{2}$,$\ldots$, $\mathcal{E}^{\text{out}}$ on $\rho_{x}^{\text{in}}$ to obtain $\rho^{1}_x, \rho^{2}_x, \ldots, \rho^{\text{out}}_x$ successively .
     \STATE \textbf{Backward:} for each training data $\{\left(\rho_{x}^{\text {in}},\tau_{x}^{\text {out}}\right)\}$, calculate $\sigma^{\text{out}} = (\tau_{x}^{\text{out}})^{1/2} ((\tau_{x}^{\text{out}})^{1/2} \rho_{x}^{\text{out}} (\tau_{x}^{\text{out}})^{1/2})^{-1/2} (\tau_{x}^{\text{out}})^{1/2}$, and then apply backward channels $\mathcal{F}^{\text{out}}$, $\mathcal{F}^{L}$,$\ldots$, $\mathcal{F}^{1}$ on $\sigma^{\text{out}}$ to successively obtain $\sigma^{L}_x, \sigma^{L-1}_x, \ldots, \sigma^{0}_x $ .
     
     \STATE \textbf{Gradients:} calculate the gradient with respect to $\theta_{(i,j),k}^{l}$ for each training data:  $\frac{\partial F_x(\rho^{\text{out}}_x, \tau^{\text{out}}_x)}{\partial \theta_{(i,j),k}^{l}} = G(\boldsymbol{\theta^{l}}, \rho^{l-1}_x, \sigma^{l}_x)$, and then take the average over the whole training dataset: $ \frac{1}{N}\sum^N_{x=1} G(\boldsymbol{\theta^{l}}, \rho^{l-1}_x, \sigma^{l}_x)$.
     \STATE \textbf{Update:} update each $\theta_{(i,j),k}^{l}$ according to $\theta_{(i,j),k}^{l} \rightarrow \theta_{(i,j),k}^{l} + \epsilon *  \frac{1}{N}\sum^N_{x=1} G(\boldsymbol{\theta^{l}}, \rho^{l-1}_x, \sigma^{l}_x)$.
\ENDFOR
\STATE Output the trained DQNN
\end{algorithmic}
\end{algorithm}

For the task of learning the ground state energy of a Hamiltonian $H$, we provide the pseudocode in Algorithm~\ref{algorithm 2}.

\begin{algorithm}[H]
\caption{Training the DQNN for learning the ground state for some Hamiltonian via the quantum backpropagation algorithm}
\label{algorithm 2}
\begin{algorithmic}
\REQUIRE The DQNN model with $L$ hidden layers, initial parameters $\boldsymbol{\theta_I}$, Hamiltonian $H$, iteration steps $s_0$ , learning rate $\epsilon$.
\ENSURE The trained DQNN
\FOR{ $s=1$ to $s_0$}
 \STATE  \textbf{Forward:} apply forward channels $\mathcal{E}^{1}$, $\mathcal{E}^{2}$,$\ldots$, $\mathcal{E}^{\text{out}}$ on initial fiducial product state $|0\cdots 0\rangle$ to obtain $\rho^{1}, \rho^{2}, \ldots, \rho^{\text{out}}$ successively .
 \STATE  \textbf{Backward:} apply backward channels $\mathcal{F}^{\text{out}}$, $\mathcal{F}^{L}$,$\ldots$, $\mathcal{F}^{1}$ on $\sigma^{\text{out}} = H$ to successively obtain $\sigma^{L}, \sigma^{L-1}, \ldots, \sigma^{0} $.
 
 \STATE \textbf{Gradients:} calculate the gradient with respect to $\theta_{(i,j),k}^{l}$:  $\frac{\partial \bar{E}(\rho^{\text{out}}, H)}{\partial \theta_{(i,j),k}^{l}} = G(\boldsymbol{\theta^{l}}, \rho^{l-1}, \sigma^{l})$  .
\STATE \textbf{Update:} update each $\theta_{(i,j),k}^{l}$ according to $\theta_{(i,j),k}^{l} \rightarrow \theta_{(i,j),k}^{l} - \epsilon * G(\boldsymbol{\theta^{l}}, \rho^{l-1}, \sigma^{l})$.
\ENDFOR
\STATE Output the trained DQNN
\end{algorithmic}
\end{algorithm}

\section*{SUPPLEMENTARY NOTE 2: Numerical results for several machine learning tasks}

In this section, we simulate the training of DQNNs by realizing the forward channels and the backward channels with matrix calculations on a classical computer, and present some numerical results.

\textbf{Task: learning a two-qubit quantum channel.} 
Here, we choose DQNN$_1$ mentioned in the main text to learn a two-qubit target quantum channel. The training dataset is the same as that in the main text.
We numerically train DQNN$_1$ with $50$ different initial parameters and show our numerical results in Supplementary Fig.~\ref{fig:2-qubit channel}. 
We observe that DQNN$_1$ shows high convergence performance, with the average converged mean fidelity above $98\%$.

We choose one learning curve (marked in triangles in Supplementary Fig.~\ref{fig:2-qubit channel}) to test the learning performance of 
DQNN$_1$. We refer DQNN$_1$ with parameters corresponding to the ending (starting) epoch of this training curve as the trained (untrained) DQNN$_1$, and then use $100$ different input quantum states to test the fidelities between their corresponding output states and the desired output states given by the target quantum channel.
As shown in the lower inset of Supplementary Fig.~\ref{fig:2-qubit channel}, for the trained DQNN$_1$, the mean fidelity exceeds $0.97$ (green bars), which separates away from the distribution of the results of the untrained DQNN$_1$ (purple bars). This contrast indicates a satisfying performance of DQNN$_1$.

\begin{figure*}[t]
\includegraphics{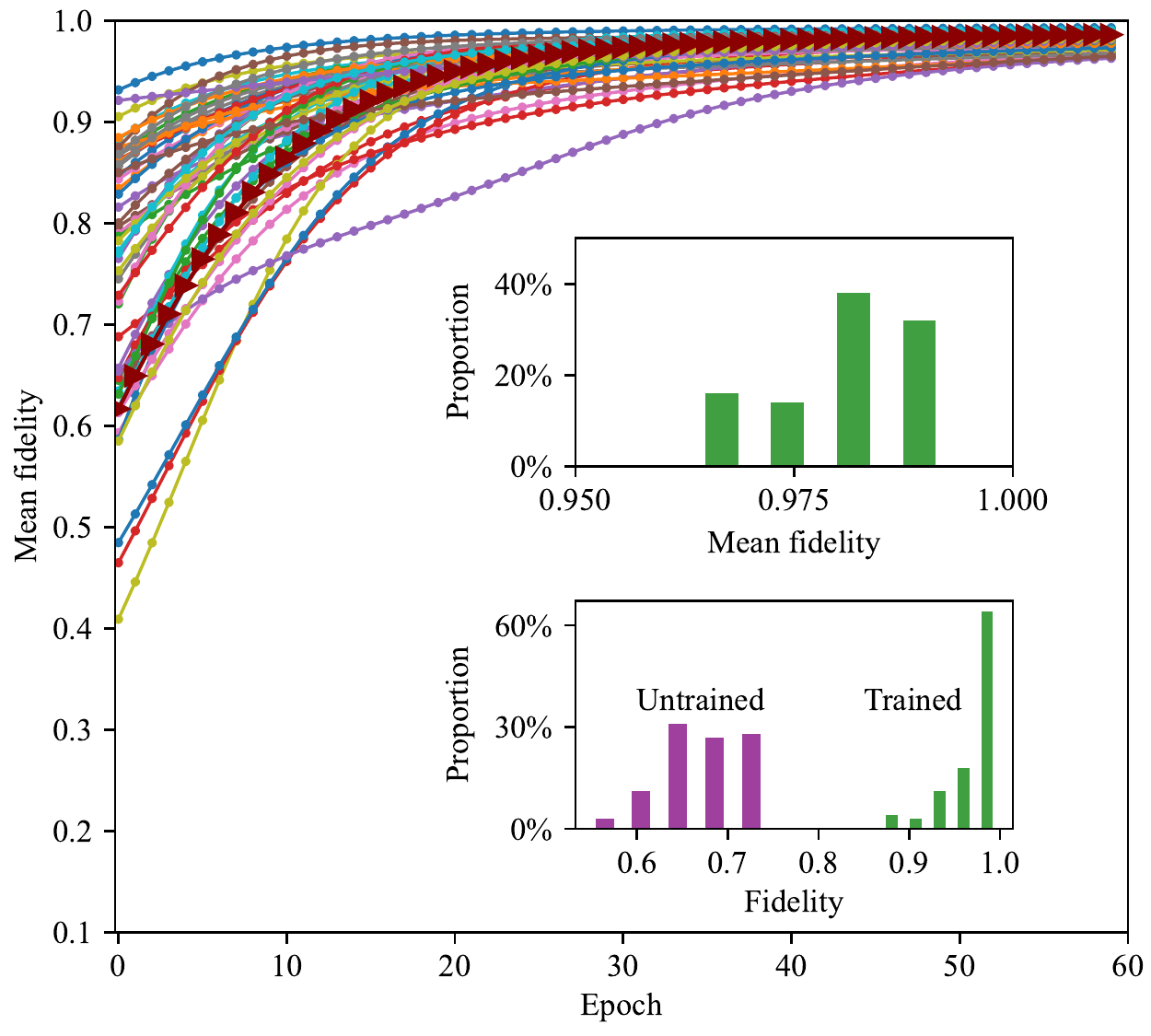}
\caption{
\textbf{Numerical results for learning a two-qubit quantum channel.}
The numerical results for training DQNN$_1$ with $50$ different initial parameters. We plot the mean fidelity as a function of the training epochs.
The upper inset shows the distribution of the converged mean fidelities for these $50$ different initial parameters.
We choose one of the learning curves (marked with triangles), and then randomly generate $100$ different input quantum states to test the fidelities between their output states given by the target quantum channel and the trained (untrained) DQNN$_1$. The results are displayed in the lower inset with the green (purple) bars showing the distribution of the fidelities for the trained (untrained) DQNN$_1$.
}
\label{fig:2-qubit channel}
\end{figure*}

\textbf{Task: learning the ground state of molecular hydrogen ($H_2$).} 
We also use DQNN$_1$ to learn the ground state energy of the molecular hydrogen Hamiltonian. We choose $50$ different initial parameters and classically simulate the training process as presented in the main text.
The results are shown in Supplementary Fig.~\ref{fig:h2}.
We observe that DQNN$_1$ converges quickly, and the average of the mean ansatz energy estimate reaches $-1.826$ (hartree) when excluding two abnormal instances with
local minima, which is very close to the theoretical value $-1.85$ (hartree). This indicates the successful application of our model.
\begin{figure*}[t]
\includegraphics{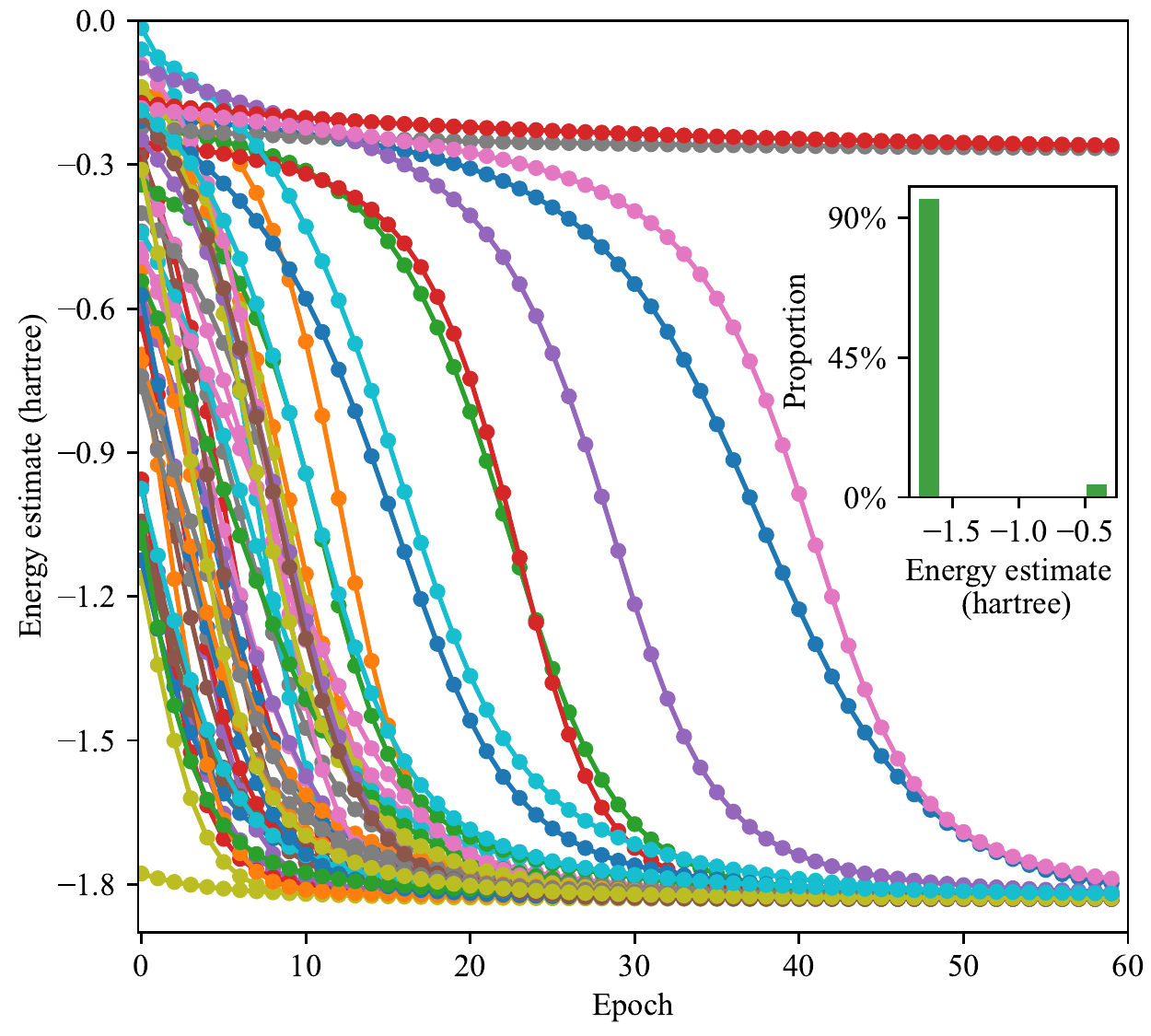}
\caption{
\textbf{Numerical results for learning the ground state energy of molecular hydrogen.} 
Energy estimate as a function of the training epochs for $50$ different initial parameters. The distribution of their converged energy estimates is displayed in the inset.
}
\label{fig:h2}
\end{figure*}

\textbf{Task: learning a one-qubit quantum channel.}
We choose DQNN$_2$ mentioned in the main text to learn a one-qubit target quantum channel. In our simulation, the training dataset is the same as in the main text.
Our numerical results for $50$ different initial parameters are summarized in Supplementary Fig.~\ref{fig:1-qubit channel}.  We observe that DQNN$_2$ shows high convergence performance in the training process, with the average converged mean fidelity above $99.5\%$.
We also choose one of these learning curves (marked in triangles in Supplementary Fig.~\ref{fig:1-qubit channel}), and refer DQNN$_2$ with parameters corresponding to the ending (starting) epoch of the training curve as the trained (untrained) DQNN$_2$.
We then use $100$ different input quantum states to test the fidelities between their corresponding output states and the desired output states given by the target quantum channel.
As shown in the lower inset of Supplementary Fig.~\ref{fig:1-qubit channel}, for the trained DQNN$_1$, the mean fidelity exceeds $0.999$ (green bars), which separates away from the distribution of the untrained DQNN$_2$ (purple bars) with the mean fidelity below $0.4$. This contrast indicates a satisfying performance of DQNN$_2$.
\begin{figure*}[t]
\includegraphics{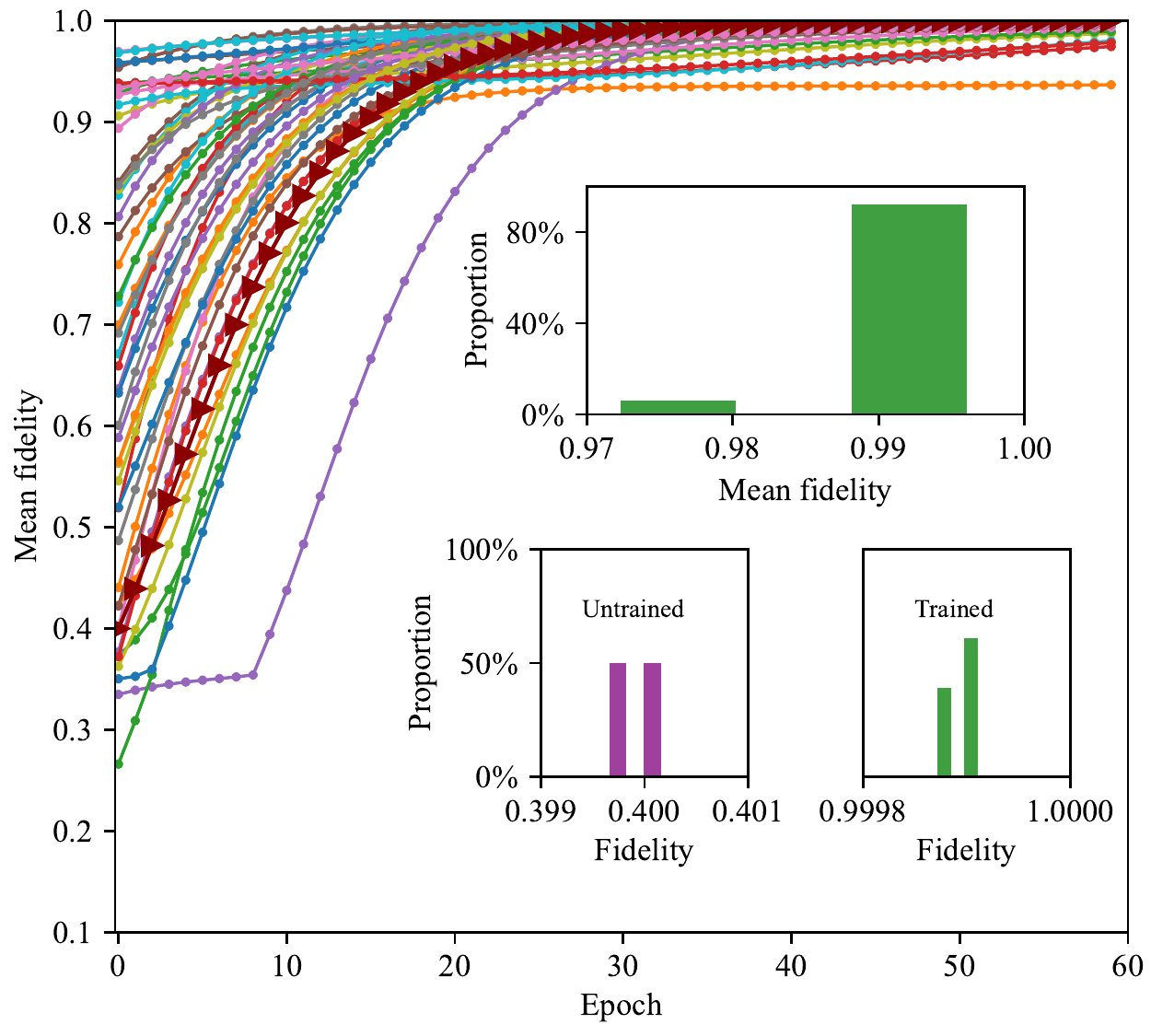}
\caption{
\textbf{Numerical results for learning a one-qubit quantum channel.} 
The mean fidelity is plotted as the function of the training epochs for $50$ different initial parameters. The distribution of their converged mean fidelities is displayed in the upper inset.
We choose one of the learning curves (marked with triangles), then randomly generate $100$ single-qubit states, separately produce their output states given by the target quantum channel and the well-trained (untrained) DQNN$_2$, and finally evaluate the corresponding fidelities between them. The distributions of the fidelities are shown in the lower inset.
}
\label{fig:1-qubit channel}
\end{figure*}

\section*{SUPPLEMENTARY NOTE 3: Experimental implementation of the DQNN}

\subsection*{Characterization of the quantum processor}
Our experiment is performed on a six-qubit superconducting quantum processor. As shown in Fig.~1(d) in the main text, the layout of qubits is carefully optimized to be a layer-by-layer structure. 
We denote these qubits as $Q_j$, where $j=1,2,...,6$, and the labels are the same as those in the main text figures. 
The detailed experimental wiring of qubit control lines and measurement lines are shown in Supplementary Fig.~\ref{fig:setup}. We summarize the characteristic parameters of our quantum processor in Table~\ref{tab:1}.

\begin{figure*}
\includegraphics{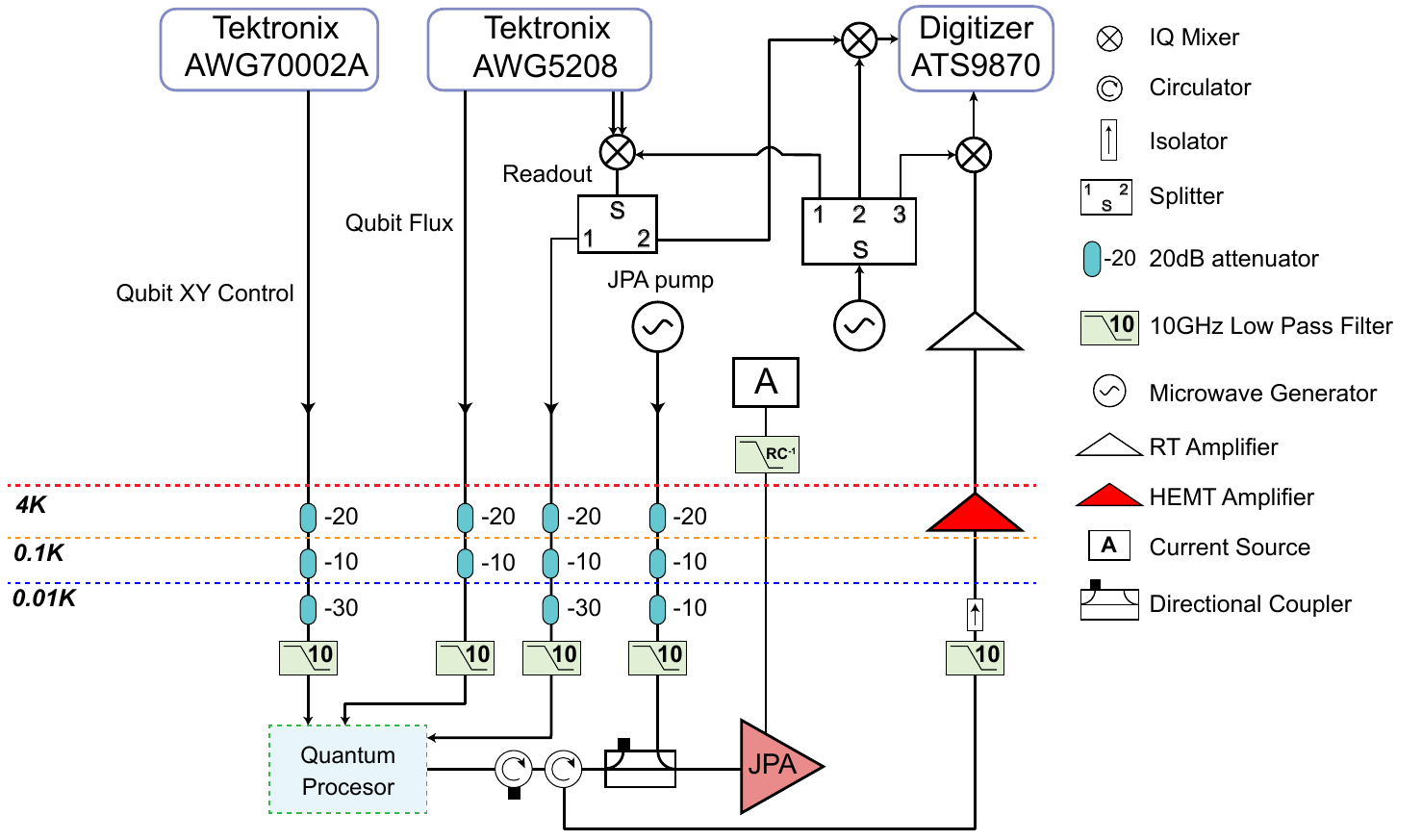}
\caption{
\textbf{The experimental wiring of qubit control lines and measurement lines.} We plot one of two identical qubit XY control lines and one of six identical qubit flux lines for simplification purpose.
}
\label{fig:setup}
\end{figure*}

\begingroup
\setlength{\tabcolsep}{12pt} 
\renewcommand{\arraystretch}{1.9} 
\begin{center}
\begin{table}
\begin{tabular}{c|c|c|c|c|c|c}
  \hline \hline
  Parameters / Qubit & $Q_1$ & $Q_2$ & $Q_3$ & $Q_4$ & $Q_5$ & $Q_6$ \\
  \hline
  Qubit working frquency $f_\text{Q}$ (GHz) & 6.413 & 6.363 & 6.328 & 6.453 & 6.083 & 6.191 \\
  \hline
  Qubit energy relaxation time $T_{1}$ ($\mu$s) & 4.2 & 6.1 & 5.1 & 8.2 & 10.0 & 10.6 \\
  \hline
  Qubit Ramsey dephasing time $T_{2}$ ($\mu$s) & 2.2 & 1.9 & 4.8 & 8.4 & 18.2 & 11.8 \\
  \hline
  Qubit anharmonicity $E_{\text{C}}/2\pi$ (MHz) & 194 & 217 & 206 & 196 & 207 & 208 \\
  \hline
  Readout resonator frequency $f_\text{R}$ (GHz) & 7.10 & 7.16 & 7.13 & 7.22 & 7.12 & 7.21 \\
  \hline
  Qubit-readout-resonator coupling strength $g_{\text{QR}}/2\pi$ (MHz) & 69 & 72 & 81 & 66 & 78 & 65 \\
  \hline
  Qubit-bus-resonator1 coupling strength $g_{\text{QB}1}/2\pi$ (MHz) & 0 & 0 & 36 & 35 & 37 & 34 \\
  \hline
  Qubit-bus-resonator2 coupling strength $g_{\text{QB}2}/2\pi$ (MHz) & 32 & 31 & 34 & 32 & 0 & 0 \\
  \hline
  Internal quality factor of the readout resonator $Q_{\text{I,R}} (10^3) $ & 102 & 83 & 130 & 15 & 85 & 92 \\\hline
  Coupled quality factor of the readout resonator $Q_{\text{C,R}} (10^3) $ & 20 & 15 & 20 & 7.6 & 7.4 & 8.0 \\
  \hline \hline
\end{tabular}
\caption{\textbf{Characteristic parameters of the quantum processor.} } 
\label{tab:1}   
\end{table}
\end{center}
\endgroup

\subsection*{Synthesize the microwave control signals}
\textbf{Timing and microwave switch control.}
We note that the microwave control signals for the single-qubit gates of $Q_1,Q_2,Q_5$ and $Q_3,Q_4,Q_6$ are directly generated by the two DAC channels of a Tektronix AWG70002A (sampling rate 25~Gs per second), respectively. The single-qubit gates are implemented by $40$~ns pulses with Gaussian envelopes. The tunability of the parameter $\theta$ in the DQNN is experimentally realized with a linear map between $\theta$ and the pulse amplitude. The individual addressing of each qubit is realized with the time domain separation of the microwave drives. In order to minimize the off-resonance crosstalk coming from the signal multiplexing, we add a fast microwave switch (activation time $<10$~ns, on-off ratio $>40$~dB) to each input XY control line, and turn on the switches only when the single-qubit gates are needed to be applied to the specific qubits. 

\textbf{Implementation of two-qubit gates with flux modulation.}
To realize the controlled-phase gate in a quantum perceptron, we adiabatically tune one of the qubit frequency to bring the $|ee\rangle$ state of the control and the target qubits into resonance with the $|gf\rangle$ state . Then the two-qubit state undergoes a periodic evolution path based on the coupling Hamiltonian of the two qubits, leaving the population of the $|ee\rangle$ state intact, but a conditional geometric phase being accumulated in the $|ee\rangle$ state. Such an operation is realized with a fast step pulse of the  external current threading the junction loop of the qubit to modify its frequency~\cite{Barends2019Diabatic}.

However, the limited bandwidth of the electronics as well as parasitic capacitances and inductances in the wiring cables lead to the distortion of the step pulses and thus the degradation of the gate performances. We adapt the method in Ref.~\cite{Roy2020Time} to mitigate this problem with the flux pulse compensation. We model the AWG response and the on-chip responses to the step pulse as low-pass filters, and apply real-time predistortions to the step pulse for compensations. The height and duration of the step pulse are optimized to minimize the errors in the swap process between $|ee\rangle$ and $|gf\rangle$. The comparison between the compensated and the uncompensated flux pulses is shown in Supplementary Fig.~\ref{fig:fluxcomp}. We note that the optimized gate parameters do not necessarily lead to a conditional $\pi$ phase, therefore, we just record the conditional phase $\phi$ and use it in the two-qubit gate in the perceptron. Moreover, the flux pulse of the target qubit will generally lead to the magnetic flux change not only in the target qubit loop, but also in other qubit loops, which causes the frequency and phase change of other qubit states. We calibrate the single-qubit phase of each qubit during the flux pulse through quantum state tomography, and compensate the flux-induced single-qubit phase in software by a phase shift of the following driving pulses. Based on the above implementation, the average fidelity of the controlled-phase gates in the DQNN is around $0.95$ for all qubit pairs.

\begin{figure*}
\includegraphics{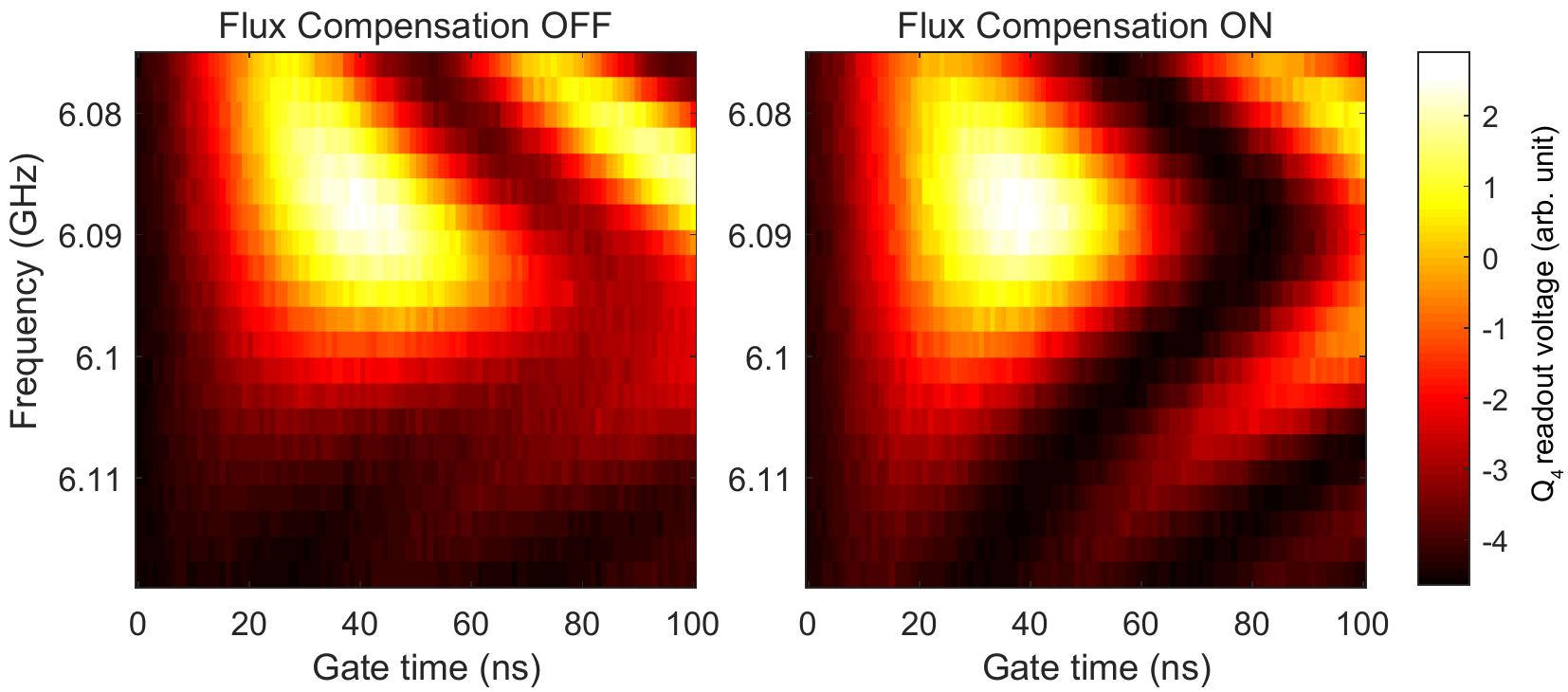}
\caption{
\textbf{The comparison between the swap operations with and without the flux compensation.} With the frequencies of all other qubits well below $5.8$~GHz, we prepare $Q_4$ and $Q_5$ in the $|eg\rangle$ state, and take a step pulse in the flux line to modulate the frequency of $Q_4$ down to reach the resonance with the $|ge\rangle$ state. The modulated qubit frequency and the duration of the step pulse are varied in the experiment. Compared with the uncompensated step pulse, the predistortion compensation method has successfully recovered the chevron pattern of the expected $|ge\rangle$ and $|eg\rangle$ swap process.
}
\label{fig:fluxcomp}
\end{figure*}

\textbf{Different working frequencies and phase compensation due to reference frame change.}
In our experiment, since the two-qubit gate requires the frequency modulation of the qubit, it is possible that the energy level resonances other than the wanted $|ee\rangle$ and $|gf\rangle$ hybridization could occur during the modulation process. Such unwanted resonances will lead to undesired state swapping that degrades the gate fidelity. In order to avoid the unwanted frequency resonances, we have set the working frequencies of the six qubits to several different configurations when executing different perceptrons in DQNN$_1$ and DQNN$_2$ (see Table~\ref{tab:U_DQNN} for details). Meanwhile, the frequency changes of the qubits require additional time-dependent phase compensation. As illustrated in Supplementary Fig.~{\ref{fig:framechange}}(a), we calibrate the time-dependent phase change of each qubit by preparing the state $(|g\rangle+|e\rangle)/\sqrt2$ with a microwave driving frequency $f_A$ when the qubit frequency is also at $f_A$, and then apply a predistorted step pulse in the flux control line to shift the qubit frequency to $f_B$. A quantum state tomography of the qubit with a driving frequency $f_B$ is performed to extract the phase accumulation caused by the frequency change. Here we fix the time interval between the tomography pulse and the state preparation pulse, and vary the time $t_{\text{shift}}$ between the state preparation pulse and the step pulse to calibrate the phase  accumulation with $t_{\text{shift}}$. The calibration result is shown in Supplementary Fig.~{\ref{fig:framechange}}(b). Such a phase accumulation is also corrected in software by shifting the phases of the driving pulses after the frequency change.

\begin{figure*}
\includegraphics{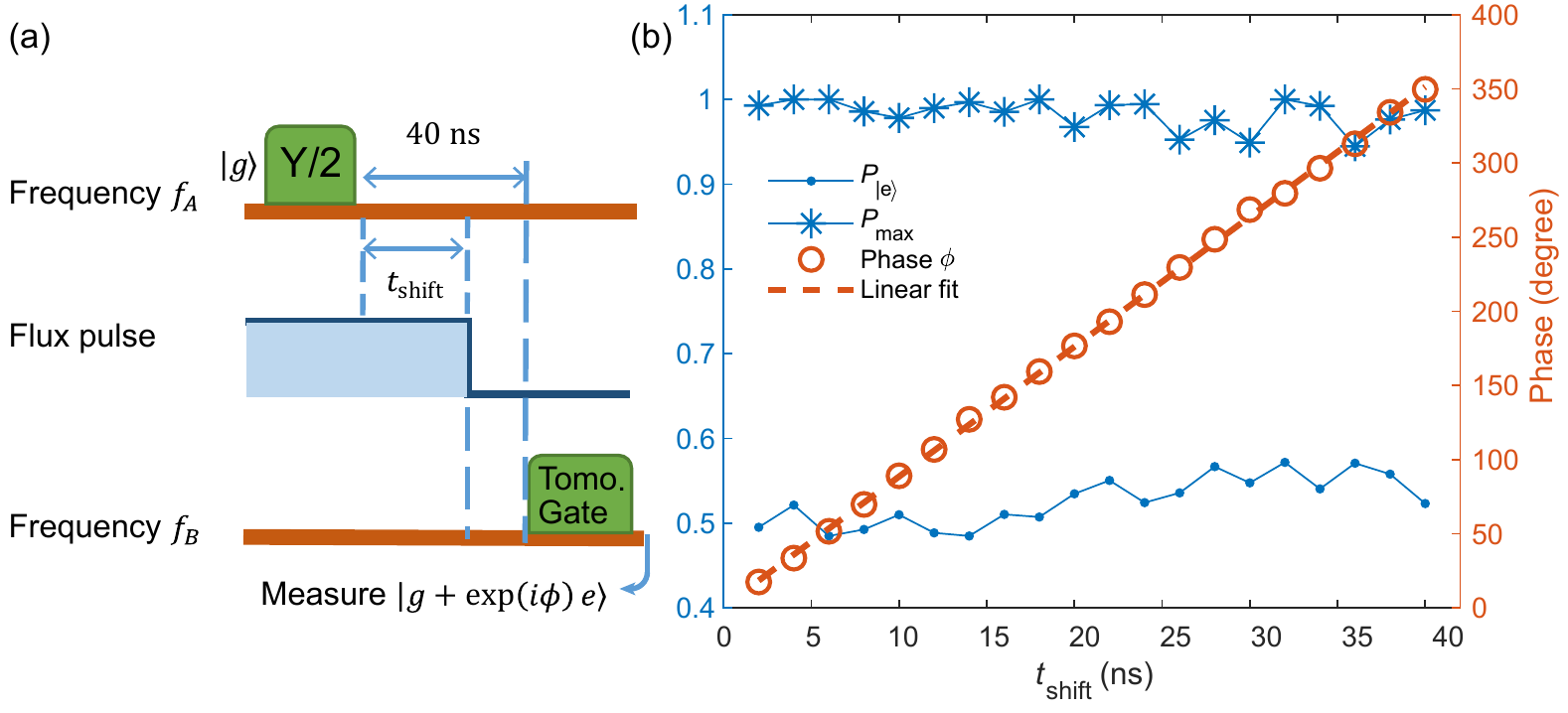}
\caption{
\textbf{Calibration of the single-qubit phase induced by the shift of the working frequency.} (a) The experimental pulse sequence. (b) The experimental result for the frequency shift process of $Q_4$. The blue dots denote the probability of measuring $|e\rangle$ state $P_{|e\rangle}$. The blue star marks denote $P_{\text{max}}$, which is the larger eigenvalue of the single-qubit density matrix. The near-unity values of $P_{\text{max}}$ indicate the measured final quantum states are close to pure states. The red circles denote the phase $\phi$ extracted from the final quantum states in the form $\ket{g}+e^{i\phi}\ket{e}$. The dashed line is a linear fit to the red circles to infer the desired frequency shift.
}
\label{fig:framechange}
\end{figure*}

\begingroup
\setlength{\tabcolsep}{7pt} 
\renewcommand{\arraystretch}{1.5} 
\begin{center}
\begin{table}
\begin{tabular}{c|c|c|c|c|c|c|c|c|c}
  \hline \hline
  \multicolumn{10}{c}{\multirow{2}{*}{DQNN$_1$}} \\   
   \multicolumn{10}{c}{\multirow{2}{*}{}} \\ \hline  \hline
   Perceptron & Qubits &$Q_1$ (GHz)& $Q_2$ (GHz)& $Q_3$ (GHz)& $Q_4$ (GHz)& $Q_5$ (GHz)& $Q_6$ (GHz)& Time (ns) & Phase  \\ \cline{1-10} 
  $U^{1}_{1,1}$ & $Q_1$, $Q_3$& \multirow{4}{*}{$6.413$} & \multirow{4}{*}{$6.364$} &  \multirow{4}{*}{$6.320$} & \multirow{4}{*}{$6.453$} & \multirow{4}{*}{$<5.8$} & \multirow{4}{*}{$<5.8$} &  62 & $175^{\circ}$\\ \cline{1-2} \cline{9-10}
  $U^{1}_{2,1}$ & $Q_2$, $Q_3$ & \multirow{4}{*}{} & \multirow{4}{*}{} & \multirow{4}{*}{} & \multirow{4}{*}{} & \multirow{4}{*}{} & \multirow{4}{*}{}  & 52 & $180^{\circ}$ \\ \cline{1-2} \cline{9-10}
  $U^{1}_{1,2}$ & $Q_1$, $Q_4$ & \multirow{4}{*}{} & \multirow{4}{*}{} & \multirow{4}{*}{} & \multirow{4}{*}{} & \multirow{4}{*}{} & \multirow{4}{*}{} & 92 & $180^{\circ}$ \\ \cline{1-2} \cline{9-10}
  $U^{1}_{2,2}$ & $Q_2$, $Q_4$ & \multirow{4}{*}{} & \multirow{4}{*}{} & \multirow{4}{*}{} & \multirow{4}{*}{} & \multirow{4}{*}{} & \multirow{4}{*}{} & 82 & $-155^{\circ}$ \\ \cline{1-10}
  $U^{2}_{1,2}$ & $Q_3$, $Q_6$ & \multirow{4}{*}{$<5.8$} & \multirow{4}{*}{$<5.8$} & \multirow{2}{*}{$6.328$} & \multirow{4}{*}{$6.453$} & \multirow{4}{*}{$6.08$} & \multirow{4}{*}{$6.191$} & $64$ & $176^{\circ}$ \\ \cline{1-2} \cline{9-10}
  $U^{2}_{1,1}$ & $Q_3$, $Q_5$ & \multirow{4}{*}{} & \multirow{4}{*}{} & \multirow{2}{*}{} & \multirow{4}{*}{} & \multirow{4}{*}{} & \multirow{4}{*}{} & 64 & $-117^{\circ}$ \\ \cline{1-2} \cline{5-5} \cline{9-10}
  $U^{2}_{2,2}$ & $Q_4$, $Q_6$ & \multirow{4}{*}{} & \multirow{4}{*}{} & \multirow{2}{*}{$<5.8$} & \multirow{4}{*}{} & \multirow{4}{*}{} & \multirow{4}{*}{} & 64 & $-157^{\circ}$ \\ \cline{1-2} \cline{9-10}
  $U^{2}_{2,1}$ & $Q_4$, $Q_5$ & \multirow{4}{*}{} & \multirow{4}{*}{} & \multirow{2}{*}{} & \multirow{4}{*}{} & \multirow{4}{*}{} & \multirow{4}{*}{} & 60 & $-165^{\circ}$ \\ \cline{1-2} \cline{9-10}
  \hline \hline
  \multicolumn{10}{c}{\multirow{2}{*}{DQNN$_2$}} \\   
  \multicolumn{10}{c}{\multirow{2}{*}{}} \\ \hline  \hline
  Pecerptron & Qubits & $Q_1$ (GHz)& $Q_2$ (GHz)& $Q_3$ (GHz)& $Q_4$ (GHz)& $Q_5$ (GHz)& $Q_6$ (GHz)& Time (ns) & Phase \\ \cline{1-10}
  $U^{1}_{1,1}$ & $Q_1$, $Q_3$ & \multirow{3}{*}{$6.413$} & \multirow{3}{*}{$6.364$} &  \multirow{3}{*}{$6.320$} & \multirow{3}{*}{$6.453$} & \multirow{3}{*}{$<5.8$} & \multirow{3}{*}{$<5.8$} &  62 & $175^{\circ}$ \\ \cline{1-2} \cline{9-10}
  $U^{2}_{1,1}$ & $Q_2$, $Q_3$ & \multirow{3}{*}{} & \multirow{3}{*}{} & \multirow{3}{*}{} & \multirow{3}{*}{} & \multirow{3}{*}{} & \multirow{3}{*}{}  & 52 & $180^{\circ}$ \\ \cline{1-2} \cline{9-10}
  $U^{3}_{1,1}$ & $Q_2$, $Q_4$ & \multirow{3}{*}{} & \multirow{3}{*}{} & \multirow{3}{*}{} & \multirow{3}{*}{} & \multirow{3}{*}{} & \multirow{3}{*}{} & 82 & $-155^{\circ}$ \\ \cline{1-10}
  $U^{4}_{1,1}$ & $Q_4$, $Q_6$ & \multirow{2}{*}{<5.8} & \multirow{2}{*}{<5.8} & \multirow{2}{*}{$<5.8$} & \multirow{2}{*}{6.453} & \multirow{2}{*}{6.08} & \multirow{2}{*}{6.191} & 64 & $-157^{\circ}$ \\ \cline{1-2} \cline{9-10}
  $U^{5}_{1,1}$ & $Q_5$, $Q_6$ & \multirow{2}{*}{} & \multirow{2}{*}{} & \multirow{2}{*}{} & \multirow{2}{*}{} & \multirow{2}{*}{} & \multirow{2}{*}{} & 60 & $-130^{\circ}$ \\ \cline{1-1} \cline{9-10}
  \hline \hline

\end{tabular}
\caption{\textbf{Experimental parameters for training DQNNs.}  In our experiments for training DQNN$_1$ and DQNN$_2$, we apply the quantum perceptrons in the order from the top to the bottom in the first column. Each perceptron acts on the qubits listed in the second column. When applying different perceptrons, we need to set the qubits to different frequencies as listed. We also show the operation time and the rotation angle for the controlled-phase gate in each perceptron.}
\label{tab:U_DQNN}   
\end{table}
\end{center}
\endgroup

\subsection*{Quantum state tomography}
We extract the quantum state $\rho^{l}$ of the qubits in each layer of the DQNN by carrying out the quantum state tomography.
To reconstruct a single-qubit state, we perform single-qubit Pauli measurements on four bases $\mathcal{S}_1=\left\{|g\rangle,|e\rangle,|+\rangle,|i\rangle\right\}$.
To reconstruct a two-qubit state,  we perform two-qubit Pauli measurements on $16$ bases $\mathcal{S}_2=\left\{ |v_1\rangle \otimes |v_2\rangle; v_1,v_2 \in  \mathcal{S}_1   \right\}$.
In our experiment, we repeat the measurement in each basis $10^4$ times to obtain a probability distribution $\vec{r}$ on the two and four computational bases for the single-qubit and two-qubit cases, respectively.
$\vec{r}$ is sent to a classical convex optimizer to find the density matrix $\rho^{l}$ that produces the distribution as close as $\vec{r}$.

\end{document}